\documentstyle[]{article}
\input amssym.def
\input amssym.tex
\newcommand{\resection}[1]{\setcounter{equation}{0}\section{#1}}

\def\Im{{\rm Im\,}
\def\Re{{\rm Re\,}}}
\newcommand{\beq}{\begin{equation}}
\newcommand{\eeq}{\end{equation}}
\newcommand{\bdm}{\begin{displaymath}}
\newcommand{\edm}{\end{displaymath}}
\newcommand{\bea}{\begin{eqnarray}}
\newcommand{\eea}{\end{eqnarray}}

\begin{document}
\setcounter{page}{0}
\topmargin 0pt
\newpage
\setcounter{page}{0}
\begin{titlepage}
\vspace{0.5cm}
\begin{flushright}{SWAT 96-97/170\\}
{???}
\end{flushright}
\begin{center}
{\Large
{Inverse scattering and the symplectic form for sine-Gordon solitons}} \\
\vspace{1cm}
{\large Edwin J. Beggs$^*$ and  Peter R. Johnson$^\dagger$
\footnote{Laboratoire de la Direction des Sciences de la Mati\`ere
du Commissariat \`a l'Energie
Atomique}}\\
\vspace{1.0cm}
{$*$ \em Department of Mathematics,\\
University of Wales Swansea,\\
Swansea,\\
SA2 8PP, Wales, UK.} \\ \&
\\
{$\dagger$ \em CEA-Saclay   \\
Service de Physique Th\'eorique  \\
F-91191 Gif-sur-Yvette Cedex \\
France}
\end{center}
\vspace{0.5cm}
\setcounter{footnote}{0}\baselineskip=12pt
\begin{center}{\bf Abstract} \\ \end{center}

\noindent
We consider the canonical symplectic form for sine-Gordon evaluated explicitly
on the solitons of the model. The integral over space in the form, which
arises because the canonical argument uses the Lagrangian density,
is done explicitly
in terms of functions arising in the
group doublecrossproduct formulation of the inverse scattering procedure,
and we are left with a simple expression given by two boundary terms.
The expression is then evaluated explicitly in terms
of the changes in the positions and momenta of the solitons, and we find
agreement with a result of Babelon and Bernard who have evaluated the
form using a different argument, where it is diagonal in terms of
`in' or `out' co-ordinates. Using the result, we also investigate the
higher conserved charges within the inverse scattering framework,
check that they Poisson commute and evaluate them on  the soliton solutions.
\end{titlepage}

\resection{Introduction}

The  purpose of this paper is to calculate the symplectic form restricted to
the
phase space
describing the dynamics of solitons in the sine-Gordon model,
in terms of the loop
groups of the inverse scattering procedure.

The aim is to be able to do Hamiltonian mechanics
with solitons, while retaining all their positions and phases.
The reader may remember that any method proceeding only via the monodromy
matrix, such as the quantum inverse scattering method \cite{standard_qism},
looses information on the positions and phases.
Some work has already been done on the soliton symplectic form restricted to
the solitons of sine-Gordon by Babelon and Bernard \cite{BB}, where a result
was obtained in terms of the positions and momenta of the solitons
by using an intuitive argument based on `in' and `out'
co-ordinates; the form becoming diagonal in these co-ordinates.
This argument solved for the form without evaluating it
explicitly (the integrals over space were not performed).

The approach adopted in this paper relies on an interesting
 simplification which occurs when the symplectic form is
expressed in the language of inverse scattering and group factorisations.
The space integral in the form can be
done explicitly by an `integration by parts trick',
and the form is then a difference of two boundary terms.
This general expression for the form (\ref{eq: the_result}) is of interest by
itself, and could be useful for applications such as quantisation -- in fact
quantisation is the real motivating force behind our construction.  However,
we go further and calculate it explicitly in terms of the changes in position
and momenta, where we indeed find agreement with the
results of Babelon and Bernard \cite{BB}.

The second purpose of the paper is to relate this to an abstract
integrable system consisting of
a group factorisation and a classical vacuum map. From this abstract point
of view we do Hamiltonian mechanics by studying Hamiltonian functions
arising as
`higher momenta' in integrable theories, and we calculate the momenta for the
sine-Gordon case. We explicitly verify a result of Olive and Turok
\cite{OT}, restricted to sine-Gordon, that the values of the higher charges
with odd Lorentz spins are zero.

\resection{Preliminaries}
We firstly define light cone coordinates on $\Bbb R^{1,1}$ by
$x_{\pm}=t\pm x$, and let
$\partial_{\pm}$ denote differentiation with respect to $x_{\pm}$
respectively.
In these coordinates the sine-Gordon equation is written as
\beq\partial_{+}\partial_{-}u=\ -{m^2\over\beta}\sin(\beta u),\label{eq:
sinegordon} \eeq
where $u(x_+,x_-)$ is a real valued function on ${\Bbb R}^{1,1}$.

    Now consider the following linear system for a function
  $$\Psi : {\Bbb R}^{1,1}\times{\Bbb C}^\ast \longrightarrow
GL_2 ({\Bbb C}) $$
\beq \partial_+\Psi (x_+,x_-,\lambda) = \Psi A(x_+,x_-,\lambda),
\quad\partial_-\Psi(x_+,x_-,\lambda) = \Psi B(x_+,x_-,\lambda). \eeq
   The linear system is overdetermined, and the compatibility condition
derived from $$\partial_{+}\partial_{-}\Psi = \partial_{-}\partial_{+}\Psi$$
is precisely
\beq \partial_{-}A -\partial_{+}B = \big[ A,B\big]. \label{eq: compat}\eeq
We shall take $A$ and $B$
of the form
\bea
A&=& -\beta(\partial_{+}u)s_3\ +\  2 m\lambda\big(\cos({\beta u\over
2})s_1 - \sin({\beta u\over 2})
s_2\big) \cr
&=&-\beta(\partial_{+}u)s_3\ +\ 2 m\lambda e^{-\beta us_3}s_{1}e^{\beta us_3}
\cr\cr
B&=&\beta(\partial_{-}u)s_3\ +\  2 m\lambda^{-1}\big(\cos({\beta
u\over 2})s_1 + \sin({\beta u\over 2})
s_2\big) \cr
&=&\beta(\partial_{-}u)s_3\ +\  2 m\lambda^{-1}e^{\beta us_3}s_{1}e^{-\beta
us_3}, \eea
where $\lambda$ can be an arbitrary complex number and is called the spectral
parameter, and
$s_1$, $s_2$, and $s_3$ are the standard anti-hermitian basis elements of
the Lie algebra
$sl(2)$
\beq
\big[s_1,s_2\big]\ =\  {1\over 2} s_3,\quad\quad
\big[s_2,s_3\big]\ =\ {1\over 2} s_1,\quad\quad
\big[s_3,s_1\big]\ =\ {1\over 2} s_2.
\eeq
Observe that the spectral parameter $\lambda$ is in the principal gradation
of the underlying loop algebra.
The compatibility conditions (\ref{eq: compat}) then immediately yield
$$(\partial_{+}\partial_{-}u+{m^2\over\beta}\sin(\beta u))s_3=0,$$
in other words $u$ is a solution to the sine-Gordon equation (\ref{eq:
sinegordon}).

\noindent For convenience we shall want to have $s_3$ diagonal, so we choose
the explicit basis
\beq
s_1\ =\ \frac14 \left(\matrix{ 0 & i \cr i & 0 }\right),  \quad\quad
s_2\ =\ \frac14 \left(\matrix{ 0 & 1 \cr -1 & 0 }\right),  \quad\quad
s_3\ =\ \frac14 \left(\matrix{ i & 0 \cr 0 & -i }\right).
\eeq
We have a simple vacuum solution to these equations given by $u$ constant with
$\cos({u \over 2})=1$. In
this case we define $J$ and $K$ by,
\beq A\ =\ J\ =\  2 m\lambda s_1,\quad {\rm and}\quad B\ =\ K\ =\  2
m\lambda^{-1} s_1.
\label{eq: JK} \eeq
The equations
$$
\partial_+\Psi_0= \Psi_0 J
,\quad{\rm and}\quad\partial_-\Psi_0= \Psi_0 K
$$
have a simple exponential solution, and
if we then subtract off this vacuum solution from the general solution
$\Psi$ by
defining
\beq \phi = \Psi_0^{-1}\Psi, \eeq
we see that $\phi$ obeys the equations
\beq \partial_+\phi = \phi A - J \phi,\quad
{\rm and}\quad\partial_-\phi=\phi B - K \phi.
\label{eq: linear} \eeq

We demand that $\lambda=0$ and $\lambda=\infty$ are regular points of the
function
$\phi(\lambda)$, by requiring that
$$\phi^{-1}\partial_+\phi = A - \phi^{-1}J \phi,\quad
{\rm and}\quad\phi^{-1}\partial_-\phi= B - \phi^{-1}K \phi$$
are free from singularities at $0$ and $\infty$,
and from this we deduce that
\beq
e^{-\beta us_3}s_{1}e^{\beta us_3}=\phi(\infty)^{-1}s_1\phi(\infty),
\label{eq: atinf} \eeq
and
\beq
 e^{\beta us_3}s_{1}e^{-\beta us_3}=\ \phi(0)^{-1}s_1\phi(0)
\label{eq: atzero}. \eeq
We can solve (\ref{eq: atinf}) by setting the normalisation of
$\phi(\lambda)$, that is $\phi(\infty)=e^{\beta u s_3}$. We can then
compute the explicit soliton solutions from $\phi(0)$ and
(\ref{eq: atzero}), which
must satisfy $\phi(0)=Ne^{-\beta u s_3}$, for $N$ a constant matrix
such that $N^{-1}s_1 N=s_1$.

To be consistent with the linear system we impose the condition, as shown in
\cite{bj1}, that
\beq U\phi(\lambda)U^{\dagger}=f(\lambda)\phi(-\lambda),\label{eq: Symmetry}
\eeq
where $f(\lambda)$ is a scalar function and $U=4s_3$.

It is well known, for example see \cite{Faddeev, Novikov}, or \cite{Edwin,bj1},
that the meromorphic solutions $\phi(\lambda)$, which are unitary on the real
axis,  can be written as products of
\beq \Bigl(P_i^\perp + {\lambda-\bar{\alpha_i}\over\lambda-\alpha_i}P_i\Bigr),
\label{eq: the_form}\eeq
where $i$ runs from one to the number of solitons in the system. Here $P_i$
is a
Hermitian projection, i.e.\ $P_i^2=P_i$, and $P_i^\dagger = P_i$. The
product can be
taken in any order, but then different $P$'s will apply.
The positions of the solitons are encoded into the $P_i$'s, and the momenta
of the $i^{\rm th}$ soliton is related in a simple way to $|\alpha_i|$, that
is the rapidity, $\theta_i$, is given by $|\alpha_i|=e^{-\theta_i}$.

The condition (\ref{eq: Symmetry})
restricts the positions of the poles $\alpha_i$ to either the imaginary
axis, which corresponds to solitons and anti-solitons (corresponding to
a single factor of the form (\ref{eq: the_form})),   or if not on the
imaginary axis, the poles come in pairs, one at $\alpha_i$ and the other
at $-\alpha_i$. This latter situation corresponds to a breather --
a bound state of a soliton and anti-soliton.

\resection{The doublecross product method.}

At this point we must beg the readers indulgence as we consider a very
abstract picture. The justifications for our digression are:

(1)\quad We will derive concrete formulae which we will apply to our
present case, the sine-Gordon model.

(2)\quad We shall find general formulae for
 the `higher conserved momenta' typical of $1+1$ dimensional integrable field
theories.

(3)\quad The expression found for the symplectic form will be based on
functions defined in this section.

Consider a group factorisation ${\cal X}={\cal GM}={\cal MG}$,
where $\cal G$ and  $\cal M$ are subgroups
of $\cal X$,
and ${\cal G}\cap M$ consists only of the identity element. We require that
any element of $\cal X$ can be uniquely factorised as $c\psi$, where
$c\in \cal M$ and
$\psi\in \cal G$, and can be uniquely factorised as $\phi f$, where
$f\in \cal M$ and
$\phi\in \cal G$. This is the definition of a group doublecross product,
and appears in the construction of Hopf algebra bicrossproducts \cite{Shahn},
 however we shall not consider  Hopf algebras in this paper.

To construct an example of an abstract integrable field theory, take a
spacetime $S$, and a function $a:S\to \cal M$, which we shall call the
classical `vacuum' map. The group ${\cal G}$ may be called the classical
phase space.
For a given classical solution $\phi_0$ in the phase space,
we perform a factorisation for all
space-time positions $s\in S$;
\beq
a(s)\, \phi_0\ =\ \phi(s)\, b(s)\ ,\quad \phi(s)\in {\cal G}\ ,\ b(s)\in
{\cal M}\ .
\label{eq: the_factorisation} \eeq
The solution to the field theory, the fields at the point $s\in S$, is
encoded into $b(s)$ and $\phi(s)$. If $S$ is a differential manifold we can
recover
a linear system from the factorisation (\ref{eq: the_factorisation}) by
differentiating it along a vector $g$ in $S$ to give $a_g\phi_0=\phi_g
b+\phi b_g$
(we use subscripts for differentiation). This can be rearranged to give
\beq
\phi_g\ =\ a_ga^{-1}\phi\ -\ \phi b_gb^{-1}\ .
\label{eq: diff_factorisation} \eeq

For the sine-Gordon model, we take $S$ to be the ordinary $1+1$ dimensional
flat  Lorentzian space-time.  The group $\cal M$ consists
 of functions from ${\Bbb C}^*$ to $GL_2({\Bbb C})$ which are
complex analytic (so that essential singularities are likely to be
present in elements of $\cal M$ at $0$ and $\infty$), and are unitary on
$\Bbb R^*$.
The group $\cal G$ consists of meromorphic
functions from ${\Bbb C}_\infty$ to $GL_2({\Bbb C})$
which are unitary on $\Bbb R_\infty$, and which satisfy the symmetry
condition (\ref{eq: Symmetry}).
The solution $\phi(\lambda)$ to the linear system (\ref{eq: linear})
is valued in ${\cal G}$, and is the same
$\phi(\lambda)$ which  appears in the factorisation
(\ref{eq: the_factorisation}). Equation (\ref{eq: diff_factorisation})
has the same form as  (\ref{eq: linear}), provided  that $a_+a^{-1}=-J$,
$a_-a^{-1}=-K$, $b_+b^{-1}=-A$, and $b_-b^{-1}=-B$. Therefore the
`vacuum' map $a(\lambda,x,t)$, in this case,
is defined as \beq a(\lambda,x,t)=e^{-Jx_+-Kx_-}\ .\label{eq: vacuum_map}\eeq

The proof of the corresponding statement for the $A_{n-1}$ affine Toda field
theories can be found in \cite{bj1}. Sine-Gordon can be recovered
from the analysis in \cite{bj1}, if we restrict ourselves to the case
$n=2$. In the more general affine Toda case the unitarity condition has 
to be dropped,
and a generalisation of (\ref{eq: Symmetry}) applies. The loss of unitarity
means that the residue $P_i$
of  the poles in each of the simple pole factors (\ref{eq: the_form}) is no
longer 
Hermitian. For sine-Gordon the
specification of the kernel of $P_i$ is fixed in terms of the image, therefore 
the problems pointed out in \cite{bj1} due to the kernel
do not apply.

\resection{Explicit calculation of some soliton solutions}
The explicit form of the soliton solutions are extremely familiar, and
have been known for quite some time, however we shall need to know
precisely how the positions of the solitons are encoded into the
residues $P_i$ and for this we need to calculate some solutions using the
method. We will also use the group factorisation (\ref{eq: the_factorisation})
to our advantage,  to simplify the calculations.

We write $\phi(\lambda)$ as a product of meromorphic loops, of the form
(\ref{eq: the_form}), which we call $\psi(\lambda)$,
times the
non-trivial normalization matrix
$$g=\phi(x,t,\infty)=e^{\beta u(x,t)s_3},$$
which we place on the right:
\beq \phi(\lambda)=\psi(\lambda).g\ . \eeq
The condition (\ref{eq: atinf}) is then automatically satisfied.
We can now calculate $u(x_+,x_-)$ from the projections $P(x_+,x_-)$ by
using the
regularity condition at $\lambda=0$ (\ref{eq: atzero})
\beq e^{2\beta us_3}s_{1}e^{-2\beta us_3}=\psi(0)^{-1}s_{1}\psi(0).
\label{eq: getsol} \eeq

It is simple to check that the unique Hermitian projection $P^{\perp}$ that
annihilates the vector $\left( \matrix{ 1 \cr \mu } \right)$ is
$$P^{\perp}=\left( \matrix {{|\mu|^2\over 1+|\mu|^2} &
{-\bar{\mu}\over 1+|\mu|^2} \cr {-\mu\over 1+|\mu|^2} & {1\over 1+|\mu|^2}
} \right). $$

\noindent{\bf The one-soliton solution}

We take one simple pole on the imaginary axis
\beq\psi(x_+,x_-,\lambda)=(P^{\perp}(x_+,x_-)+{\lambda+i\kappa\over\lambda-i
\kappa}P(x_+,x_-))\eeq
where $\kappa$ is real and positive. The symmetry condition
(\ref{eq: Symmetry}) implies that
$$U(P^{\perp}+{\lambda+i\kappa\over\lambda-i\kappa}P)
U^\dagger =f(\lambda)(P^{\perp}+{\lambda-i\kappa\over\lambda+i\kappa}P)$$
$$({\lambda+i\kappa\over\lambda-i\kappa})(UPU^{\dagger}+{\lambda-i\kappa\over
\lambda+i\kappa}
UP^{\perp}U^{\dagger})=f(\lambda)(P^{\perp}+{\lambda-i\kappa\over\lambda+i\kappa
}
P).$$
We see that
$$f(\lambda)=({\lambda-i\kappa\over\lambda+i\kappa}),$$
and
\beq UPU^{\dagger}=P^{\perp}=1-P. \label{eq: newsym}\eeq
The group factorisation (\ref{eq: the_factorisation}) implies that the
image of $P$ projects onto the space
$$V(x_+,x_-)=e^{2im(\kappa x_+-\kappa^{-1}x_-)s_1}V_0\ ,$$ where $V_0$ is an
arbitrary initial space. $V_0$ must be one dimensional for there to be
non-trivial solutions. Alternatively $P^\perp$ is the unique projection which
annihilates this space.

We take $V_0$ to be spanned by the vector
\beq \left( \matrix{ 1 \cr 1  } \right)(iQ)^{-1/2}+
\left( \matrix{ 1 \cr -1  } \right)(iQ)^{1/2}, \label{eq: Q} \eeq
where $Q$ is real. Then $V(x_+,x_-)$ is spanned by the vector
\bea
\left( \matrix{ 1 \cr 1  } \right)(iW)^{-1/2}+
\left( \matrix{ 1 \cr -1  } \right)(iW)^{1/2}
\ =\  \left( \matrix{ (iW)^{-1/2}+(iW)^{1/2} \cr (iW)^{-1/2}-(iW)^{1/2} }
\right),
\eea
where
$$W=Qe^{m(\kappa x_+-\kappa^{-1}x_-)},$$
and thus the hermitian projection $P^{\perp}(x_+,x_-)$ which annihilates
$V(x_+,x_-)$ is
\beq P^\perp(x_+,x_-)=\frac12\left( \matrix { 1 & -({1+iW\over 1-iW}) \cr
-({1-iW\over 1+iW}) & 1 } \right)
\label{eq: onesol} \eeq
This projection is also consistent with the symmetry condition
(\ref{eq: Symmetry})  which has now been
translated into the form (\ref{eq: newsym}). Any other
choice of initial space $V$ would not have satisfied this condition, and
indeed it is this choice which enforces the reality of the solution.

We have
$$\psi(x_+,x_-,0)=(1-2P)=\left( \matrix{ 0 & -({1+iW\over 1-iW}) \cr
-({1-iW\over 1+iW}) & 0 } \right).$$
Now
\bea
\frac i4\pmatrix{0 \cr 1}e^{-i\beta u}&=&(1-2P)s_1(1-2P)\pmatrix{1 \cr 0} \cr
&=&-\Bigl({1-iW\over 1+iW}\Bigr)(1-2P)s_1\pmatrix{0 \cr 1} \cr
&=&\Bigl({1-iW\over 1+iW}\Bigr)^2\frac i4\pmatrix{0 \cr 1},
\eea
so the one-soliton solution is
\beq
e^{-i{\beta u\over 2}}={1-iW\over 1+iW}.\label{eq: one_soll}\eeq
We note that $\log |Q|$ is essentially the position $x_0$ of this
soliton (after suitable normalisation) at some fixed time,
since $x_0$ is given through $Q$ by
\beq {\rm sign}(Q) W=
e^{m((\kappa-\kappa^{-1})t + (\kappa+\kappa^{-1})(x-x_0)}, \eeq
so
\beq\log(|Q|)=-m(\kappa+\kappa^{-1})x_0\ .\label{eq: x_0}\eeq
Now $Q\in{\Bbb R}$, and the choice $Q<0$, $Q>0$ corresponds to taking solitons
or anti-solitons. Note that we can also change the sign of $\kappa$ to move
from a soliton to an anti-soliton.

The position of the pole at $i\kappa$ is related to the
rapidity $\theta$ of the soliton by $|\kappa|=e^{-\theta}$.

\noindent{\bf The two-soliton solution}

We take
$$\psi(x_+,x_-,\lambda)=(P_{1}^{\perp}+{\lambda+i\kappa_1\over\lambda-i\kappa_1}
P_{1})(P_{2}^{\perp}+{\lambda+i\kappa_2\over\lambda-i\kappa_2}P_{2}),$$
for $\kappa_1$ and $\kappa_2$ real and positive.
We can choose to write the unitary meromorphic loop corresponding to the pole
at $i\kappa_{2}$ to the left in the product:
\beq
(P_{1}^{\perp}+{\lambda+i\kappa_1\over\lambda-i\kappa_1}P_{1})(P_{2}^{\perp}+
{\lambda+i\kappa_2\over\lambda-i\kappa_2}P_{2})=
(P_{3}^{\perp}+{\lambda+i\kappa_2\over\lambda-i\kappa_2}P_{3})(P_{4}^{\perp}
+{\lambda+i\kappa_1\over\lambda-i\kappa_1}P_{4}). \label{eq: reorder} \eeq
The projections $P_{3}$ and $P_{4}$ are uniquely determined. We do this
because we cannot immediately solve for $P_{2}(x_+,x_-)$ since it is not
ordered to the left, but on the other hand we can of course solve for
$P_{3}(x_+,x_-)$.

Imposing the symmetry condition (\ref{eq: Symmetry}) implies
$$U(P_{1}^{\perp}+{\lambda+i\kappa_{1}\over\lambda-i\kappa_{1}}
P_{1})U^{\dagger}U(P_{2}^{\perp}+{\lambda+i\kappa_{2}
\over\lambda-i\kappa_{2}}P_{2})U^{\dagger}
$$
$$ =f(\lambda)(P_{1}^{\perp}+{\lambda-i\kappa_{1}\over\lambda+i\kappa_{1}}
P_{1})(P_{2}^{\perp}+{\lambda-i\kappa_{2}\over\lambda+
i\kappa_{2}}P_{2}), $$
and hence,
$$({\lambda+i\kappa_1\over\lambda-i\kappa_1})({\lambda+i\kappa_2\over\lambda-
i\kappa_2})(UP_1U^{\dagger}+{\lambda-i\kappa_1\over\lambda+i\kappa_1}
UP_1^{\perp}U^{\dagger})
(UP_2U^{\dagger}+{\lambda-i\kappa_2\over\lambda+i\kappa_2}
UP_2^{\perp}U^{\dagger})
$$
$$=f(\lambda)(P_1^{\perp}+{\lambda-i\kappa_1\over\lambda+i\kappa_1}P_1)
(P_2^{\perp}+{\lambda-i\kappa_2\over\lambda+i\kappa_2}P_2).
$$
It follows that
$$ U P_1U^\dagger =P^{\perp}_1 \qquad U P_2U^\dagger =P^{\perp}_2, $$
repeating the argument for the simple pole factors reversed implies
$$ UP_3U^\dagger =P_3^\perp \qquad U P_4U^{\dagger}=P_4^\perp. $$

Only the conditions for $P_1$ and $P_3$ are strictly needed for the
calculation of $P_2$. Indeed we can take the one-soliton solution
(\ref{eq: onesol})
for $P_1$ and $P_3$ separately.
\beq
P_1^\perp=\frac12\left( \matrix { 1 & -({1+iW_1\over 1-iW_1}) \cr
-({1-iW_1\over 1+iW_1}) & 1 } \right) \eeq
and
$$V_{3}(x_+,x_-)=\left( \matrix{ (iW_2)^{-1/2}+(iW_2)^{1/2} \cr
(iW_2)^{-1/2}-(iW_2)^{1/2} } \right) $$
where
\beq W_{1}=Q_{1}e^{m(\kappa_{1}x_+-\kappa_1^{-1}x_-)} \qquad {\rm and}\qquad
W_{2}=Q_{2}e^{m(\kappa_{2}x_+-\kappa_2^{-1}x_-)}. \label{eq: Q1Q2}\eeq

We rewrite equation (\ref{eq: reorder}) as
$$U_{1}(\lambda)((P_{2}^{\perp}+{\lambda+i\kappa_2\over\lambda-i\kappa_2}P_{2})
=(P_{3}^{\perp}+{\lambda+i\kappa_2\over\lambda-i\kappa_2}P_{3})U_{4}(\lambda)$$
evaluating the residue at $\lambda=i\kappa_2$
$$U_{1}(i\kappa_2)P_{2}=P_{3}U_{4}(i\kappa_2)$$
$$P_{2}=U_{1}(i\kappa_2)^{-1}P_{3}U_{4}(i\kappa_2).$$
$P_{2}$ projects onto the image of $U_{1}(i\kappa_2)^{-1}P_{3}U_{4}(i\kappa_2)$
 and this is the image of $U_{1}(i\kappa_2)^{-1}P_{3}$ since
$U_{4}(i\kappa_2)$ is invertible. This space is the space annihilated by
$P_{2}^{\perp}$. $P_{2}$ can now be computed explicitly. So we have
$$ V_{2}(x_+,x_-)=U_{1}(i\kappa_2)^{-1}V_{3}(x_+,x_-),$$
and it follows that
$$V_2=\pmatrix{{1+W_1W_2-iY^{-1}W_1+iY^{-1}W_2\over Y^{-1}(1-iW_1)}\cr
{1+W_1W_2+iY^{-1}W_1-iY^{-1}W_2\over Y^{-1}(1+iW_1)}}$$
and
\beq
P_{2}=\left( \matrix{ \frac12 & \frac12{(1+iW_1)(1+W_1W_2-iY^{-1}W_1+
iY^{-1}W_2)\over
(1-iW_1)(1+W_1W_2+iY^{-1}W_1-iY^{-1}W_2)} \cr
\frac12{(1-iW_1)(1+W_1W_2+iY^{-1}W_1-iY^{-1}W_2)\over(1+iW_1)(1+W_1W_2-iY^{-1}W_
1+iY^{-1}W_2)}
& \frac12 } \right) \label{eq: 2solexplicit} \eeq
where
$$Y=({\kappa_{2}-\kappa_{1}\over \kappa_{2}+\kappa_{1}}).$$
We then find that
$$\psi(x_+,x_-,0)=(1-2P_{1})(1-2P_{2})=\pmatrix
{{1+iY^{-1}W_1-iY^{-1}W_2+W_1W_2\over
1-iY^{-1}W_1+iY^{-1}W_2+W_1W_2} & 0 \cr
0 & {1-iY^{-1}W_1+iY^{-1}W_2+W_1W_2\over1+iY^{-1}W_1-iY^{-1}W_2+W_1W_2}}. $$
Now $$
\frac i4\pmatrix{0 \cr 1}e^{-i\beta
u}=((1-2P_1)(1-2P_2))^{-1}s_1(1-2P_1)(1-2P_2)
\pmatrix{1 \cr 0}, $$
so we get
\beq
e^{-i\beta u}=\Biggl({(1+iY^{-1}W_{1}-iY^{-1}W_{2}+W_{1}W_{2})\over
(1-iY^{-1}W_{1}+iY^{-1}W_{2}+W_{1}W_{2})}\Biggr)^2\ . \label{eq: soweget}\eeq
If we adjust the initial subspaces $V_{1}(0,0)$ and $V_{3}(0,0)$ by
\beq W_{1}\rightarrow -YW_{1}, \qquad W_{2}\rightarrow
YW_{2}, \label{eq: adjust} \eeq
and let \beq X=Y^{2}=\biggl({\kappa_{2}-\kappa_{1}\over \kappa_{2}+
\kappa_{1}}\biggr)^{2}, \eeq
then we get the familiar two-soliton solution:
\beq e^{-i\beta {u\over 2}}={(1-iW_{1}-iW_{2}-XW_{1}W_{2})\over
(1+iW_{1}+iW_{2}-
XW_{1}W_{2})}.\label{eq: two_sol_sol}
\eeq
The solution (\ref{eq: soweget}) is written in `left-most' ordered 
co-ordinates,
that is coordinates
given by
taking the $Q$'s defined by the projection 
for each pole in the left-most ordered position in the
factorisation. 
We note that the adjustment (\ref{eq: adjust}) to these co-ordinates
is a subtle re-normalisation
of the standard $Q$'s, which enter the standard form of the
solution (\ref{eq: two_sol_sol}). This
adjustment will be crucial for us in obtaining the final version of the
symplectic form written in terms of the positions and momenta of the solitons.

\noindent{\bf The breather solution}

We choose the poles to be at $\alpha$ and $-\alpha$, where $\alpha$ is not
on the imaginary axis.
\bea
\psi(x_+,x_-,\lambda)&=&(P_{1}^{\perp}+{\lambda-\bar{\alpha}\over\lambda-
\alpha}P_{1})(P_{2}^{\perp}+{\lambda+\bar{\alpha}\over\lambda+\alpha}P_{2}) \cr
&=&(P_{3}^{\perp}+{\lambda+\bar{\alpha}\over\lambda+
\alpha}P_{3})(P_{4}^{\perp}+{\lambda-\bar{\alpha}\over\lambda-\alpha}P_{4}).
\label{eq: brpsi}
\eea
The symmetry condition (\ref{eq: Symmetry}) implies that
\beq U P_1U^\dagger =P_3, \qquad U P_2U^\dagger =P_4  \label{eq:
newsymmetry2}. \eeq
This time we take for $Q\in {\Bbb C}$,
\beq V_1=Q^{-1/2}\pmatrix{ 1
\cr 1} + Q^{1/2}\pmatrix{ -1 \cr 1}
\label{eq: phase} \eeq
and then, writing $W(\alpha)=e^{m(\alpha x_+ + \alpha^{-1}x_-)}$,
$$V_1(x_+,x_-)=\pmatrix{ 1 \cr 1}(QW(\alpha))^{-1/2} +
\pmatrix{ -1 \cr 1}(QW(\alpha))^{1/2},$$
splitting $QW(\alpha)$ into a pure phase and modulus, $QW(\alpha)=UW$, we have
\bea
V_1(x_+,x_-)&=&\pmatrix{ 1 \cr 1}(UW)^{-1/2} + \pmatrix{ -1 \cr 1}(UW)^{1/2}
\cr \cr
&=&\left( \matrix{ (UW)^{-1/2}-(UW)^{1/2} \cr (UW)^{-1/2}+(UW)^{1/2} }
\right),
\eea
where, setting $\alpha=v+ik$,
$$
U=e^{im(vx_++{vx_-\over v^2+k^2})}=e^{im((v+{v\over v^2+k^2})(t-t_0)+(-v+
{v\over v^2+k^2})(x-x_0))} $$
and
$$
W=e^{m(-kx_++{kx_-\over v^2+k^2})}=e^{m((-k+{k\over v^2+k^2})(t-t_0)+(k+
{k\over v^2+k^2})(x-x_0))}. $$
Here $x_0$ and $t_0$ are fixed through the complex valued $Q$.
We can write any $Q\in {\Bbb C-\{0\}}$ as
$Q=e^{m(\alpha q_1 + \alpha^{-1}q_2)}$, since $\alpha$ has a non-zero
imaginary part, so that $\alpha$ and $\alpha^{-1}$ span the complex plane,
and where $q_1,q_2\in{\Bbb R}$.
Then, on remembering that $x_\pm = t\pm x$, $x_0$ and $t_0$ are
related to $q_1$ and $q_2$ by
$$q_1=-(x_0+t_0),\qquad q_2=-t_0+x_0.$$

Then
$$
P_1=\frac12\pmatrix{{(U-W)(1-UW)\over U(1+W^2)} & -{(-1+UW)(W+U)\over
U(1+W^2)} \cr
-{(-U+W)(UW+1)\over U(1+W^2)} & {(U+W)(1+UW)\over U(1+W^2)} },
$$
and to be consistent with Eqn. (\ref{eq: newsymmetry2}),
 we must start with the
vector $V_3(0,0)$:
$$V_3(0,0)=Q^{1/2}\pmatrix{ 1
\cr 1} + Q^{-1/2}\pmatrix{ -1 \cr 1},$$
and we get the evolution  (recall that the pole is at $\lambda=-\alpha$)
\beq V_3(x_+,x_-)=\left( \matrix{ -((UW)^{-1/2}-(UW)^{1/2}) \cr (UW)^{-1/2}+
(UW)^{1/2} } \right). \label{eq: V3}\eeq
Using the same arguments as for the two soliton case we find that, putting
$s=k/v$,
$$
V_{2}=\pmatrix{ -{(isW+isU^2W-U-UW^2)\over U(1+W^2)} \cr
{(UW+1)(isW+isU^2W+U+UW^2)\over (-1+UW)U(1+W^2)} },
$$
and
$$
P_2=\pmatrix{ -\frac12{(-1+UW)(U-W)\over U(1+W^2)} &
-\frac12 {(W+U)(isW+isU^2W-U-UW^2)(-1+UW)\over(isW+isU^2W+U+UW^2)U(1+W^2)} \cr
\frac12 {(U-W)(isW+isU^2W+U+UW^2)(1+UW)\over(isW+isU^2W-U-UW^2)U(1+W^2)} &
\frac12{(1+UW)(U+W)\over U(1+W^2)} }, $$
so
\bea
\psi(0)&=&(1-{2is\over 1+is}P_1)(1-{2is\over 1+is}P_2) \cr \cr
&=&\pmatrix{ -{(s^2+1)(isW+isU^2W+U+UW^2)\over(isW+isU^2W-U-UW^2)(1+is)^2} & 0
\cr
0 & -{(s^2+1)(isW+isU^2W-U-UW^2)\over(isW+isU^2W+U+UW^2)(1+is)^2} }. \eea
The result follows:
\beq
e^{{-i\beta u\over 2}}=\Bigl({isU^{-1}+isU+W^{-1}+W\over
isU^{-1}+isU-W^{-1}-W}\Bigr).\label{eq: classical_breather}
\eeq

\resection{The symplectic form}

We have seen that the loop $\phi_0$ specifies a solution to the sine Gordon
equation for all space-time. We can consider the set of such $\phi_0$
to be the phase space of the sine Gordon system. Then a change $v_0$ in
$\phi_0$ would represent a change in $u$ at all values in
space-time. Given two such changes $v_0$ and $w_0$ we should be able to
calculate the
canonical symplectic form (\ref{eq: form}), and this is what we shall do in
this section. Here we denote the time derivative of $u$ by $\dot{u}$.

\par\noindent{\bf Lemma $5.1$}

With $b$ defined by the factorisation (\ref{eq: the_factorisation})
\beq
(b_x b^{-1})_v=\beta (\delta_v \dot{u})s_3 +2m\beta(\delta_v
u)(\lambda^{-1}e^{\beta u s_3}[s_3,s_1]e^{-\beta u s_3} +
\lambda e^{-\beta u s_3}[s_3,s_1]e^{\beta u s_3}) \label{eq: lemma1}
\eeq

\noindent{\bf Proof}  Using the results for $b$ just prior to  Eqn.
 (\ref{eq: vacuum_map})
we can write
\bea b_x b^{-1}&=&b_+b^{-1}-b_-b^{-1}=B-A\cr\cr
&=&\beta(\partial_- u+\partial_+ u)s_3 + 2m (\lambda^{-1}e^{\beta u
s_3}s_1e^{-\beta u s_3}-\lambda e^{-\beta u
s_3}s_1e^{\beta u s_3})\cr\cr
&=&\beta \dot{u} s_3 + 2m (\lambda^{-1}e^{\beta u
s_3}s_1e^{-\beta u s_3}-\lambda e^{-\beta u
s_3}s_1e^{\beta u s_3}).\eea
Now vary $u$ by a parameter we refer to as $v$. We obtain
$$(b_x b^{-1})_v=\beta (\delta_v \dot{u}) s_3 +2m\beta(\delta_v
u)(\lambda^{-1}e^{\beta u s_3}[s_3,s_1]e^{-\beta u s_3} +
\lambda e^{-\beta u s_3}[s_3,s_1]e^{\beta u s_3}). \quad\square$$

\noindent The canonical symplectic form \cite{BB}, derived from the Lagrangian
density formulation of the sine-Gordon model, is
$$
\omega\ =\ {\beta^2\over 4}\int_{-\infty}^\infty \delta u\wedge
\delta{\dot{u}}\, dx.
$$
This can be written as
\beq
\omega(v,w)\ =\ {\beta^2\over 4}\int_{-\infty}^\infty\Bigl(\delta_v u
\delta_w\dot{u}-\delta_w u\delta_v \dot{u}\Bigr)\, dx\label{eq: form}
\eeq
in terms of variations $v=\delta_v \phi$ and $w=\delta_w\phi$.

\noindent{\bf Proposition $5.2$}

The form (\ref{eq: form}) is given by the formula
\beq \omega(v,w)={\frac{1}{2\pi i}}\int_\gamma {d\lambda\over\lambda}
{\rm Trace\
}\int_{-\infty}^\infty \Bigl(b_xb^{-1}\Bigr)_v \phi^{-1}w\,dx
\label{eq: sympform}\eeq
where the contour $\gamma$ is a small clockwise circle
and a large anti-clockwise circle around the origin. Here `small' means
that all the poles of
the meromorphic function $\phi$ lie outside the contour, and `large' means
that all the poles
lie inside it.

\noindent{\bf Proof}

\noindent Using Lemma $5.1$,
$$\omega(v,w)={\frac{1}{2\pi i}}\int_\gamma {d\lambda\over\lambda}
{\rm Trace\ }\int_{-\infty}^\infty\Bigl(\beta(\delta_v \dot{u})s_3
\qquad\qquad\qquad\qquad$$
\beq \qquad\qquad
+2m\beta(\delta_v u)(\lambda^{-1}e^{\beta u s_3}[s_3,s_1]e^{-\beta u s_3}
+\lambda e^{-\beta u s_3}[s_3,s_1]e^{\beta u
s_3})\Bigr)\phi^{-1}w\,dx.\eeq
We calculate this integral in two parts, for the first part
$$
{\frac{1}{2\pi i}}
\beta\int_\gamma{d\lambda\over\lambda}{\rm Trace\
}\int_{-\infty}^\infty
(\delta_v\dot{u})s_3\phi^{-1}w\, dx.$$
The ${\frac{1}{2\pi i}}\int_\gamma {d\lambda\over\lambda}$ is equivalent to
evaluating
at $\lambda\rightarrow\infty$ minus evaluating at $\lambda=0$, that is
$$\beta\int_{-\infty}^\infty
(\delta_v\dot{u}){\rm Trace\
}\Bigl(s_3\phi^{-1}w(\infty)-s_3\phi^{-1}w(0)\Bigr)\, dx,$$
recall that $\phi(\infty)=e^{\beta u s_3}$ and $\phi(0)=Ne^{-\beta u
s_3}$, where $N$ is some constant normalisation matrix. Then, we get
\bea &&\beta^2\int_{-\infty}^\infty
(\delta_v\dot{u}){\rm Trace\ }(s_3^2)(\delta_w u+\delta_w u)\, dx\cr
&=&-{\beta^2\over 4}\int_{-\infty}^\infty
(\delta_v\dot{u})(\delta_w u)\, dx.\eea

The second term is
$${2m\beta\over 2\pi i}\int_\gamma {d\lambda\over\lambda}{\rm Trace\
}\int_{-\infty}^\infty
(\delta_v u)\Bigl(\lambda^{-1}e^{\beta u s_3}[s_3,s_1]e^{-\beta u s_3}
+ \lambda e^{-\beta u s_3}[s_3,s_1]e^{\beta u s_3}\Bigr)\phi^{-1}w\,
dx, $$
and integrating around the small circle in $\gamma$ gives the $\lambda^0$
coefficient of the expansion about $\lambda=0$ of
$$-2m\beta\int_{-\infty}^\infty(\delta_v u)\lambda^{-1}{\rm Trace\ }
s_3[e^{\beta u s_3}s_1e^{-\beta u s_3},\phi^{-1}w]\, dx$$
$$=-2m\beta\int_{-\infty}^\infty(\delta_v u){\rm Trace\ }
s_3[e^{\beta u s_3}s_1e^{-\beta u s_3},(\phi^{-1}w)'(0)]\, dx.$$
To calculate this, consider the derivative of $a\phi_0=\phi b$ in the
direction $x_-$;
$$\lambda\phi^{-1}\phi_-=\lambda\beta(\partial_- u)s_3+2me^{\beta u
s_3}s_1 e^{-\beta u s_3} - 2m \phi^{-1}s_1\phi,$$
and now apply $\frac{d}{d\lambda}$ at $\lambda=0$ to get
$$\phi^{-1}\phi_-(0)=\beta
(\partial_-u)s_3-2m(\phi^{-1}s_1\phi'(0)-\phi^{-1}\phi'\phi^{-1}s_1\phi(0))\ ,$$
(we use prime to denote $\frac{d}{d\lambda}$)
and apply $\phi^{-1}\phi_-(0)=-\beta (\partial_- u)s_3$,
$$-\beta (\partial_- u)s_3=-m(\phi^{-1}s_1\phi'(0)-
\phi^{-1}\phi'\phi^{-1}s_1\phi(0)).$$
Conjugating this with $\phi(0)$, and noting that
$\phi(0)s_3\phi^{-1}(0)=Ns_3N^{-1}$, we have
\bea
-\beta (\partial_- u)Ns_3N^{-1}&=&m(\phi'\phi^{-1}(0)s_1 -
s_1\phi'\phi^{-1}(0)) \cr
&=&m[\phi'\phi^{-1}(0),s_1]. \eea
Now vary this in the direction $w$
$$-\beta(\partial_-
u_w)Ns_3N^{-1}=m[w'\phi^{-1}(0)-\phi'\phi^{-1}w\phi^{-1}(0),s_1]\ ,$$
and conjugate by $\phi(0)$ again get
\bea -\beta
(\partial_-u_w)s_3&=&m[\phi^{-1}w'(0)-\phi^{-1}\phi'\phi^{-1}w(0),
\phi^{-1}(0)s_1\phi(0)]\cr\cr
&=&m[(\phi^{-1}w)'(0),e^{\beta u s_3}s_1e^{-\beta u s_3}].\eea
This gives the integral of the second term around the small circle in
$\gamma$ as
$$\frac{\beta^2}4\int_{-\infty}^\infty(\delta_v u)(\delta_w(\partial_-
u))\, dx.$$
Now we turn to the integral around the large circle in $\gamma$. This
is the $\lambda^0$ part of the Laurent expansion of
$$2m\beta \int_{-\infty}^\infty (\partial_v u)\lambda {\rm Trace\
}s_3[e^{-\beta u s_3}s_1 e^{\beta u s_3},\phi^{-1}w]\, dx$$
about $\lambda=\infty$. We shall calculate this from the $x_+$
equation of the linear system:
$$\phi^{-1}\phi_+=-\beta (\partial_+u)s_3+2m\lambda e^{-\beta u s_3}s_1
e^{\beta u s_3}- 2m \lambda \phi^{-1}s_1\phi,$$
taking the limit as $\lambda\rightarrow\infty$, and using
$\phi^{-1}\phi_+(\infty)=\beta(\partial_+u)s_3$,
$$\beta(\partial_+u)s_3=m\lim_{\lambda\rightarrow\infty}\lambda(e^{-\beta
u s_3}s_1e^{\beta u s_3}-\phi^{-1}s_1\phi),$$
if we set $\phi'(\infty)=\lim_{\lambda\rightarrow\infty}\lambda
(\phi(\lambda)-\phi(\infty))$, then
$$\beta(\partial_+u)s_3=m(\phi^{-1}\phi'\phi^{-1}s_1\phi(\infty)-\phi^{-1}s_1
\phi'(\infty)),$$
since $\phi(\infty)$ commutes with $s_3$, we can conjugate to get
$$\beta(\partial_+u)s_3=m[\phi'\phi^{-1}(\infty),s_1],$$
and differentiating along the direction $w$ to get
$$\beta(\partial_+u_w)s_3=m[w'\phi^{-1}(\infty)-
\phi'\phi^{-1}w\phi^{-1}(\infty),s_1],$$
conjugating again
\bea \beta(\partial_+u_w)s_3&=&m[\phi^{-1}w'(\infty)-
\phi^{-1}\phi'\phi^{-1}w(\infty),\phi^{-1}s_1\phi(\infty)]\cr\cr
&=&m[(\phi^{-1}w)'(\infty),e^{-\beta u s_3}s_1e^{\beta u s_3}].\eea
Then the contribution to the second term for the large circle in
$\gamma$ is
$${\beta^2\over 4}\int_{-\infty}^\infty(\delta_v
u)(\delta_w(\partial_+u))\, dx\ ,$$
so the total contribution from the second term is
\bea {\beta^2\over 4}\int_{-\infty}^\infty(\delta_v
u)(\delta_w(\partial_+u + \partial_- u))\ =\
{\beta^2\over 4}\int_{-\infty}^\infty(\delta_v u)(\delta_w\dot{u})\, dx\ ,\eea
and the total is the required formula
$$\omega(v,w)={\beta^2\over 4}\int_{-\infty}^\infty \Bigl(\delta_v
u\delta_w\dot{u} -
\delta_w u\delta_v\dot{u}\Bigr)\, dx\ . \quad \square
$$

Now we are left with the problem of how to perform the $x$ integral in
\beq \omega(v,w)={\frac{1}{2\pi i}}\int_\gamma {d\lambda\over\lambda}
{\rm Trace\
}\int_{-\infty}^\infty\Bigl(b_xb^{-1}\Bigr)_v\phi^{-1}w\,dx
\label{eq: sympform1}\ ,\eeq
and for this
we shall need the following Lemma:

\noindent {\bf Lemma $5.3$}
$$- \bigl(\phi^{-1}w\bigr)_x+
[b_xb^{-1},\phi^{-1}w]=(b_xb^{-1})_w$$

\noindent{\bf Proof}

Begin by varying the equation $a\phi_0=\phi b$
by the parameter $w_0$, a change in $\phi_0$, and find
\beq aw_0=wb+\phi b_w,\label{eq: v0} \eeq
where $w$ is the corresponding change in $\phi$. On
differentiating with respect to $x$ we get
$$a_xw_0=w_xb+wb_x+\phi_x b_w+\phi b_{wx}.$$
Now substitute for $w_0$ from (\ref{eq: v0}) and for $\phi_x$ from
(\ref{eq: diff_factorisation})
$$a_xa^{-1}wb+a_x a^{-1}\phi b_w=w_xb + wb_x+ a_xa^{-1}\phi b_w-\phi
b_x b^{-1}b_w+\phi b_{wx},$$
rearrange,
$$a_x a^{-1}w -w_x-wb_xb^{-1}=\phi(b_{wx}b^{-1}-b_xb^{-1}b_w b^{-1}),$$
and substitute for $a_x a^{-1}$ from (\ref{eq: diff_factorisation})
$$\phi_x\phi^{-1}w+\phi b_xb^{-1}\phi^{-1}w-w_x - wb_x b^{-1}=\phi(b_x
b^{-1})_w\ .$$
Multiplying by $\phi^{-1}$ on the left and rearranging gives the
required result
$$-(\phi^{-1} w)_x+[b_xb^{-1},\phi^{-1}w]=(b_x b^{-1})_w\ .\quad\square $$

\noindent{\bf Proposition $5.4$}
\beq\omega(v,w)={1\over 2\pi i}\int_\gamma{d\lambda\over\lambda}{\rm Trace\ }
\Bigl[b_v b^{-1}\phi^{-1}w\Bigr]_{x=-\infty}^{x=\infty}\label{eq: the_result}
\eeq

\noindent{\bf Proof}

We can write
\beq \omega(v,w)\ =\ {\frac{1}{2\pi i}}\int_\gamma {d\lambda\over\lambda}
{\rm Trace\
}\int_{-\infty}^\infty\Bigl\{\bigl(b_vb^{-1}\bigr)_x\phi^{-1}w
+ [b_vb^{-1},b_x b^{-1}]\phi^{-1}w\Bigr\}\,dx\ . \eeq
Now integrate by parts and reorder the commutation relation using the trace
property:
$$\omega(v,w)={\frac{1}{2\pi i}}\int_\gamma {d\lambda\over\lambda}
{\rm Trace\ }\Bigl\{\Big[b_vb^{-1}\phi^{-1}w\Big]_{x=-\infty}^{x=\infty}
+\int_{-\infty}^\infty dx\, b_v b^{-1}\Big(- \bigl(\phi^{-1}w\bigr)_x+
[b_xb^{-1},\phi^{-1}w]\Big)\Bigr\}$$
Using Lemma $5.3$, we can write
$$\omega(v,w)={\frac{1}{2\pi i}}\int_\gamma {d\lambda\over\lambda}
{\rm Trace\ }\Bigl\{\Big[b_vb^{-1}\phi^{-1}w\Big]_{x=-\infty}^{x=\infty}
+\int_{-\infty}^\infty dx\, b_v b^{-1}(b_x b^{-1})_w\Bigr\}\ ,$$
and we observe  that the  integral around $\gamma$  of
$b_v b^{-1}(b_x b^{-1})_w$
vanishes since the function is analytic between the circles comprising
$\gamma$.$\quad\square$

\resection{The abstract symplectic form.} It might be thought that
the form of the symplectic form we derived in the last section would be
highly dependent on the structure of the sine-Gordon equation. However
the formula
 (\ref{eq: the_result}) actually gives a closed 2-form associated to a group
doublecross product under very general conditions, as we now show.

Let ${\cal X}={\cal GM}={\cal MG}$ be a group doublecross product
with an adjoint invariant inner product $\big<,\big>$
on its Lie algebra.
 From the factorisation
$a\phi_0=\phi b$ and a change $(\phi_0;v_0)$ in $\phi_0$ we define
derivatives
$v=\phi_v=D_{(\phi_0;v_0)}\phi$ and $b_v=D_{(\phi_0;v_0)}b$ (in these
derivatives
$a$ is kept constant).

Our aim is to define a closed 2-form over the phase space $\cal G$ with
coordinate $\phi_0$. To do this we first define another 2-form
$\tau_a$ on $\cal G$, for a given $a\in \cal M$, as
$$
\tau_a(\phi_0;v_0,w_0)\ =\ \big<\, b_v b^{-1}\, ,\, \phi^{-1} w \, \big>\ -\
\big<\, b_w b^{-1}\, ,\, \phi^{-1} v \, \big>\ .
$$

\noindent{\bf Proposition $6.1$}
\begin{eqnarray*}
d\tau_a(\phi_0;y_0,v_0,w_0)& = &
\big<\, \big[ w_0\phi_0^{-1},y_0\phi_0^{-1}\big]\, ,\, v_0\phi_0^{-1} \, \big>
\ -\ \big<\, \big[\phi^{-1}w,\phi^{-1}y\big]\, ,\, \phi^{-1}v \, \big>\\
& & \ -\ \big<\, \big[b_w b^{-1},b_y b^{-1}\big]\, ,\, b_v b^{-1} \, \big>\ .
\end{eqnarray*}

\noindent{\bf Proof}

First calculate the derivative
\begin{eqnarray*}
\tau_a'(\phi_0;v_0,w_0;y_0)& = &
\big<\, b_{vy} b^{-1}-b_v b^{-1}b_y b^{-1}\, ,\, \phi^{-1} w \, \big> +
 \big<\, b_v b^{-1}\, ,\, \phi^{-1} \phi_{wy}-\phi^{-1} y\phi^{-1} w \,
\big> \\
&&-\big<\, b_{wy} b^{-1}-b_w b^{-1}b_y b^{-1}\, ,\, \phi^{-1} v \, \big>-
\big<\, b_w b^{-1}\, ,\, \phi^{-1} \phi_{vy}-\phi^{-1} y\phi^{-1} v \, \big>\ .
\end{eqnarray*}
By adding the other two terms and using symmetry of the double derivative,
we find
\begin{eqnarray*}
d\tau_a(\phi_0;y_0,v_0,w_0) &=&
\tau_a'(\phi_0;v_0,w_0;y_0)\ -\
\tau_a'(\phi_0;y_0,w_0;v_0)\ +\
\tau_a'(\phi_0;y_0,v_0;w_0)   \\
&=&\big<\, \big[b_y b^{-1},b_v b^{-1}\big]\, ,\, \phi^{-1} w \, \big>\ +\
\big<\, \big[b_w b^{-1},b_y b^{-1}\big]\, ,\, \phi^{-1} v \, \big>  \\
& &\ +\
\big<\, \big[b_v b^{-1},b_w b^{-1}\big]\, ,\, \phi^{-1} y \, \big>\ +\
  \big<\, b_v b^{-1}\, ,\, \big[\phi^{-1} w,\phi^{-1} y\big] \, \big>   \\
& &\ +\
\big<\, b_y b^{-1}\, ,\, \big[\phi^{-1} v,\phi^{-1} w\big] \, \big>\ +\
\big<\, b_w b^{-1}\, ,\, \big[\phi^{-1} y,\phi^{-1} v\big] \, \big>\ .
\end{eqnarray*}
Now we try to simplify this expression, starting with the factorisation
$a \phi_0 = \phi b$,
and differentiating in the direction $(\phi_0;v_0)$ to get
$a v_0 = v b  + \phi b_v$,
or in a more useful form
$$
\phi^{-1}a v_0b^{-1}\ =\ b_v b^{-1}\ +\ \phi^{-1}v\ .
$$
Now using this result successively,
\begin{eqnarray*}
\big<\, \big[b_w b^{-1},b_y b^{-1}\big]\, ,\, \phi^{-1} v \, \big>&=&
\big<\, \big[b_w b^{-1},b_y b^{-1}\big]\, ,\, \phi^{-1}a v_0b^{-1} \,
\big>\ -\
\big<\, \big[b_w b^{-1},b_y b^{-1}\big]\, ,\, b_v b^{-1} \, \big>   \\
&=&
\big<\, \big[b_w b^{-1},\phi^{-1}a y_0b^{-1}\big]\, ,\, \phi^{-1}a
v_0b^{-1} \, \big>\ -\
\big<\, \big[b_w b^{-1},\phi^{-1}y\big]\, ,\, \phi^{-1}a v_0b^{-1} \, \big>  \\
& &\ -\ \big<\, \big[b_w b^{-1},b_y b^{-1}\big]\, ,\, b_v b^{-1} \, \big> \\
&=&
\big<\, \big[\phi^{-1}a w_0b^{-1},\phi^{-1}a y_0b^{-1}\big]\, ,\,
\phi^{-1}a v_0b^{-1} \, \big>  \\
& & -\
\big<\, \big[\phi^{-1}w,\phi^{-1}a y_0b^{-1}\big]\, ,\, \phi^{-1}a
v_0b^{-1} \, \big>\\
& &  -\
\big<\, \big[b_w b^{-1},\phi^{-1}y\big]\, ,\, \phi^{-1}a v_0b^{-1} \, \big>
\ -\  \big<\, \big[b_w b^{-1},b_y b^{-1}\big]\, ,\, b_v b^{-1} \, \big>  \\
&=&
\big<\, \big[ w_0\phi_0^{-1},y_0\phi_0^{-1}\big]\, ,\, v_0\phi_0^{-1} \,
\big>\ -\
\big<\, \big[\phi^{-1}w,b_y b^{-1}\big]\, ,\, \phi^{-1}a v_0b^{-1} \, \big>\\
&\ & -\
\big<\, \big[\phi^{-1}w,\phi^{-1}y\big]\, ,\, \phi^{-1}a v_0b^{-1} \, \big>
\ -\
\big<\, \big[b_w b^{-1},\phi^{-1}y\big]\, ,\, \phi^{-1}a v_0b^{-1} \, \big> \\
& &  -\ \big<\, \big[b_w b^{-1},b_y b^{-1}\big]\, ,\, b_v b^{-1} \, \big> \\
&=&
\big<\, \big[ w_0\phi_0^{-1},y_0\phi_0^{-1}\big]\, ,\, v_0\phi_0^{-1} \, \big>
\ -\
\big<\, \big[\phi^{-1}w,b_y b^{-1}\big]\, ,\, b_v b^{-1} \, \big> \\
& & -\
\big<\, \big[\phi^{-1}w,b_y b^{-1}\big]\, ,\, \phi^{-1}v \, \big>
\ -\
\big<\, \big[\phi^{-1}w,\phi^{-1}y\big]\, ,\, b_v b^{-1} \, \big>\\
& & -\
\big<\, \big[\phi^{-1}w,\phi^{-1}y\big]\, ,\, \phi^{-1}v \, \big>
 \ -\
\big<\, \big[b_w b^{-1},\phi^{-1}y\big]\, ,\, b_v b^{-1} \, \big> \\
 & & -\
\big<\, \big[b_w b^{-1},\phi^{-1}y\big]\, ,\, \phi^{-1}v \, \big>
\ -\  \big<\, \big[b_w b^{-1},b_y b^{-1}\big]\, ,\, b_v b^{-1} \, \big> \ .
\end{eqnarray*}
Substitution of this into the formula above for $d\tau_a$, and using
adjoint invariance of the
inner product, gives the result.\quad $\square$

\smallskip
Now we define a 2-form $\omega$ on $\cal G$ by the formula
$$
\omega(\phi_0;v_0,w_0)\ =\ \lim_{R\to\infty}\Big[\big<\, b_v b^{-1}\, ,\,
\phi^{-1} w \, \big>\ -\
\big<\, b_w b^{-1}\, ,\, \phi^{-1} v \, \big>\Big]_{x=-R}^R\ ,
$$
where we remember that $a$ is a function on space-time.
This is the difference in the values of $\tau_a$ between two points in
space-time
lying either side of the `interesting' region, that is where the fields are
substantially different from the vacuum. If the fields are not compactly
supported, but merely
decreasing, we take a limit `$x\to\pm\infty$', as the points tend to
regions where the
fields take more vacuum-like values.
But what does `vacuum' mean in our group picture? A brief look at the
asymptotic form of the
 soliton solutions
will give a possible answer. For the sine-Gordon or principal chiral model the
 function $\phi(t,x):\Bbb C_\infty\to GL_n$ tends to a limit matrix valued
function commuting
 with $J$ and $K$
as $x\to\pm\infty$, c.f. Lemma $10.1$, below. The subgroup consisting of such
functions is abelian. In general, we assume that $\phi(s)$ tends to
a limiting value in an abelian subgroup,
which we call ${\cal G}_\infty\subset \cal G$.

\noindent{\bf Proposition $6.2$}

Suppose that $\phi(s)$ in the factorisation $a(s)\phi_0=\phi(s)b(s)$ tends
to a limiting value in the abelian subgroup ${\cal G}_\infty\subset\cal G$
as $x\to\infty$
and as $x\to-\infty$. Also suppose that the inner product $\big<,\big>$
vanishes on the
Lie algebra of the group $\cal M$. Then
 $d\omega=0$.

\noindent{\bf Proof}

This is more or less direct from the previous proposition. The
$\big<\, \big[ w_0\phi_0^{-1},y_0\phi_0^{-1}\big]\, ,\, v_0\phi_0^{-1} \,
\big>$
term in $d\tau_a$ cancels on taking differences. The term
$\big<\, \big[b_w b^{-1},b_y b^{-1}\big]\, ,\, b_v b^{-1} \, \big>$
is zero since the inner product vanishes on $ m$. The term
$\big<\, \big[\phi^{-1}w,\phi^{-1}y\big]\, ,\, \phi^{-1}v \, \big>$
tends to zero as $x\to\pm\infty$ since the commutator tends to zero by the
abelian subgroup condition.$\quad\square$

\resection{The abstract higher momenta.}

Having decided that $d\omega=0$ under certain realistic conditions, we
can try to
calculate values of $\omega$ on certain vectors $(\phi_0;w_0)$. But which
vectors to use?
There is a certain easy choice of vector, which will lead to
the higher conserved momenta for sine-Gordon.

\noindent{\bf Proposition $7.1$}

Consider $c$ in the Lie algebra of $\cal M$ which commutes with $J$ and $K$,
that is it commutes with $s_1$. Then $c$ brings about a change in $\phi_0$
through
the factorisation $e^{cr}\phi_0=\phi_0(r)m$, for some $m\in{\cal M}$.
The corresponding infinitesimal change $w_0$ of $\phi_0$ is given by the
formula
$c\phi_0=w_0+\phi_0 d$, where $d$ is also in the Lie algebra of
$\cal M$.
Then the vector field
$(\phi_0;w_0)$ is a Hamiltonian flow and is generated by the Hamiltonian
\beq
f_c(\phi_0)\ =\ 2
\lim_{R\to\infty}\Big[\big<\, \vartheta(\phi)\phi^{-1}\, ,\,  h \,
\big>\Big]_{x=-R}^R  \ ,\label{eq: Ham}
\eeq
where $\vartheta$ is a derivation operator acting on the Lie algebra of
$\cal M$,
defined by Eqn. (\ref{eq: derivation}),
 and  $h$ is an element in the Lie algebra of $\cal M$ such that $\vartheta
h=c$.
This means that
\beq \omega(\phi_0;v_0,w_0)\  =\ D_{(\phi_0;v_0)}f_c(\phi_0)\ .
\eeq

\noindent{\bf Proof}

 If we insert the formula for $w_0$ into
$aw_0=wb+\phi b_w$, we find
\beq
\phi^{-1}aca^{-1}\phi\ -\ bdb^{-1}\ =\ \phi^{-1}w\ +\ b_w b^{-1}\ .
\eeq
Using our assumption that
the inner product is zero on $ m$, we obtain
\begin{eqnarray}
\big<\, b_v b^{-1}\, ,\, \phi^{-1} w \, \big> & = &
\big<\, b_v b^{-1}\, ,\, \phi^{-1}aca^{-1}\phi \, \big>  \cr
& = &
\big<\, \phi b_v b^{-1}\phi^{-1}\, ,\, aca^{-1} \, \big>  \cr
&=&
\big<\, a v_0\phi_0^{-1}a^{-1}-v\phi^{-1}\, ,\, aca^{-1} \, \big>  \cr
&=&
\big<\,  v_0\phi_0^{-1}\, ,\, c \, \big>\ -\
\big<\, v\phi^{-1}\, ,\, aca^{-1} \, \big>  \label{eq: inner1}\\
\big<\, b_w b^{-1}\, ,\, \phi^{-1} v \, \big>\ &=&
\big<\, \phi^{-1} aca^{-1} \phi-bdb^{-1}-\phi^{-1}w\, ,\, \phi^{-1} v \,
\big> \cr
 &=&
\big<\,  aca^{-1}\, ,\,  v\phi^{-1} \, \big> \ -\
\big<\, bdb^{-1}\, ,\, \phi^{-1} v \, \big>\ -\
\big<\, \phi^{-1}w\, ,\, \phi^{-1} v \, \big>  \cr
 &=&
\big<\,  aca^{-1}\, ,\,  v\phi^{-1} \, \big> \ -\
\big<\, d\, ,\, \phi_0^{-1} v_0 \, \big>\ -\
\big<\, \phi^{-1}w\, ,\, \phi^{-1} v \, \big>  \ .\label{eq: inner2}
\end{eqnarray}
We now subtract Eqns. (\ref{eq: inner1}) and (\ref{eq: inner2})
and calculate $\omega(\phi_0;v_0,w_0)$, for some arbitrary change $v_0$.
 On taking differences between $x=\pm\infty$,
 we see that the
$\big<\, d\, ,\, \phi_0^{-1} v_0 \, \big>$ and
$\big<\,  v_0\phi_0^{-1}\, ,\, c \, \big>$ terms vanish
as $c$ and $d$ are independent of $a$, so we are left with
\beq
\omega(\phi_0;v_0,w_0)\ =\
\lim_{R\to\infty}\Big[
\big<\, \phi^{-1}w\, ,\, \phi^{-1} v \, \big>\ -\ 2\big<\, v\phi^{-1}\, ,\
 aca^{-1} \, \big>\Big]_{x=-R}^R\ .
\label{eq: ome}\eeq
In what follows, we restrict ourselves to the case
where $c$ commutes with every $a$,
and with the limit subgroup ${\cal G}_\infty\subset {\cal G}$. Then
$\phi^{-1}c \phi-bdb^{-1}=\phi^{-1}w+ b_wb^{-1}$,
where $\phi^{-1}c \phi\to c$ as $x\to\pm\infty$. This means that
$\phi^{-1}w=\pi_{ g}(\phi^{-1}c \phi\ -\ bdb^{-1})\to 0$
as $x\to\pm\infty$ ($\pi_{ g}$ is the projection to the
Lie algebra of $\cal G$).  Then the first term of (\ref{eq: ome}) vanishes and
we can rewrite Eqn. (\ref{eq: ome}) as
\beq
\omega(\phi_0;v_0,w_0)\ =\ -2
\lim_{R\to\infty}\Big[
\big<\, v\phi^{-1}\, ,\, c \, \big>\Big]_{x=-R}^R\ .
\label{eq: ome2} \eeq
This might be a sufficiently simple formula to calculate
$\omega(\phi_0;v_0,w_0)$
for various vectors $(\phi_0;v_0)$. However it is our purpose to do Hamiltonian
mechanics, so we would like to answer the question `is there a Hamiltonian
function
giving rise to the vector field $(\phi_0;w_0)$ generated by $c\in m$?'. To
do this
we would have to show that $\omega(\phi_0;v_0,w_0)$ was the derivative in
the direction $(\phi_0;v_0)$
of a function of $\phi_0$ and $c$.

Suppose that there is a 1-parameter automorphism $\Theta:\Bbb R\times X\to X$
which preserves the subgroups $\cal G$ and $\cal M$, and the inner product
on the Lie
algebra of $X$.
Then there is a derivation\footnote{Loosely speaking, in the applications,
$\vartheta$ can be thought of as $\vartheta=\lambda
\frac{d}{d\lambda}$} $\vartheta$ on the Lie algebra defined by
\beq\vartheta(y)=\Theta''(0,e;0,y;1,0)\label{eq: derivation} \eeq
 ($e$ is the group identity), and this
preserves the inner product,
that is $\big<\vartheta y,z\big>+\big< y,\vartheta z\big>=0$.

If there is an $h$ in the Lie algebra of $\cal M$ so that $\vartheta h=c$, then
\beq
\omega(\phi_0;v_0,w_0)\ =\ -2
\lim_{R\to\infty}\Big[
\big<\, v\phi^{-1}\, ,\, \vartheta h \, \big>\Big]_{x=-R}^R
\ =\ 2
\lim_{R\to\infty}\Big[
\big<\, \vartheta(v\phi^{-1})\, ,\,  h \, \big>\Big]_{x=-R}^R
\ .
\label{eq: ifthereis}\eeq
For our purposes we wish to to swap the order of $\vartheta$ and
$D_{(\phi_0;v_0)}$ in this expression. To do this we proceed cautiously and
apply $\vartheta$ to $v\phi^{-1}$ to get
$\vartheta(v)\phi^{-1}-v\phi^{-1}\phi_\vartheta\phi^{-1}$, where
$\vartheta(v)=\Theta''(0,\phi;0,v;1,0)$ and
$\phi_\vartheta=\Theta'(0,\phi;1,0)$.
From the last formula and the symmetry of double derivatives
we see that $D_{(\phi_0;v_0)}\phi_\vartheta=\vartheta(v)$, so
\begin{eqnarray*}\vartheta(v\phi^{-1})& =&
D_{(\phi_0;v_0)}\big(\phi_\vartheta \phi^{-1}\big)\ -\
\big[v\phi^{-1},\phi_\vartheta\phi^{-1}\big]\ , \\
\big<\, \vartheta(v\phi^{-1})\, ,\,  h \, \big>& = &
D_{(\phi_0;v_0)}\big<\, \phi_\vartheta \phi^{-1}\, ,\,  h \, \big>\ -\
\big<\, \big[v\phi^{-1},\phi_\vartheta\phi^{-1}\big]\, ,\,  h \, \big>\ .
\end{eqnarray*}
If $\vartheta$ preserves the abelian subgroup ${\cal G}_\infty$ then the
Lie bracket
$\big[\phi_{\vartheta}\phi^{-1}\, ,\, v\phi^{-1}\big]$ tends to zero
as $x\to\pm\infty$, and
we can write Eqn. (\ref{eq: ifthereis})
$$
\omega(\phi_0;v_0,w_0)\  =\ D_{(\phi_0;v_0)}\Big(2
\lim_{R\to\infty}\Big[
\big<\, \phi_\vartheta \phi^{-1}\, ,\,  h \, \big>\Big]_{x=-R}^R \Big)
\ ,
$$
or in other words the vector field $(\phi_0;w_0)$
 generated by an element $c\in m$
is a Hamiltonian flow, with Hamiltonian
$$
f_c(\phi_0)\ =\ 2
\lim_{R\to\infty}\Big[\big<\, \vartheta(\phi)\phi^{-1}\, ,\,  h \,
\big>\Big]_{x=-R}^R  \qquad\square
$$

Now that we have expressions for
Hamiltonian functions $f_c$ we should show that these functions Poisson
commute.

\noindent{\bf Proposition $7.2$}

The Hamiltonians $\{f_c\}$ Poisson commute.

\noindent{\bf Proof}

Suppose that the vector $(\phi_0;v_0)$ is generated by another element
$\hat c$ in the Lie algebra of $\cal M$ which commutes with $s_1$. Then
$\hat c\phi_0=v_0+\phi_0 \hat
d$, where $\hat d$ is also in
the Lie algebra of $\cal M$. Eqn. (\ref{eq: ome2}) applied for an arbitrary
$v_0$,  we can insert  this particular $v_0$ to find,
$$
\omega(\phi_0;v_0,w_0)\
=\ 2
\lim_{R\to\infty}\Big[
\big<\, v\phi^{-1}\, ,\,  c \, \big>\Big]_{x=-R}^R \ =\ -2
\lim_{R\to\infty}\Big[
\big<\, \phi^{-1}\hat c\phi\, ,\,  c \, \big>\Big]_{x=-R}^R\ =\ 0
\ .
$$
This means that the function $f_c$ Poisson commutes with $f_{\hat{c}}$, and
therefore the $\{f_c\}$  are in involution. $\quad\square$

As we shall see, in the case
of the sine-Gordon equation, we can take $c=J$ or $c=K$ to obtain  the
Hamiltonians  representing the total energy and
momentum in light-cone co-ordinates. The higher momenta are obtained
by taking the infinite number of other possible $c$'s.
 Since the energy is the Hamiltonian generator of time
translations, and the
momentum is the generator of space translations, all the higher momenta are
conserved by these
translations. The 1-parameter flow corresponding to the Hamiltonian $f_c$
on the phase space is
given by the factorisation $(r,\phi_0)\mapsto \phi_0(r)$,
where $\phi_0(r)\in \cal G$ is the solution to the factorisation problem
$e^{rc}\phi_0=\phi_0(r)d$
for some $d\in \cal M$.

\resection{The higher momenta for sine-Gordon.}
Here we shall specialise the results of the last section to the solitons in
the sine-Gordon model.
The 1-parameter automorphism of the loop group $X$ is
given by $\Theta(s,\rho)(\lambda)=\rho(s\lambda)$ for $\lambda\in\Bbb C^*$,
which
is actually the Lorentz boost for the system, giving $\vartheta=
\lambda\frac{d}{d\lambda}$. The inner product is
$$
\big<y,z\big>\ =\ \frac{1}{2\pi i}\  {\rm Trace}\ \int_\gamma\,
\frac{d\lambda}{\lambda}y(\lambda) z(\lambda)\ ,
$$
which is $\Theta$ invariant. We can choose $h(\lambda)=s_1\lambda^n$,
giving $c(\lambda)=ns_1\lambda^n$.
The meromorphic loops for sine-Gordon split into two cases,
solitons and breathers, and we shall calculate the higher momenta for both
cases.

The loop $\phi$ for a single soliton  is
$\phi=N\psi e^{\beta u s_3}$, where $N$ commutes with $s_1$, and
$$\psi=\Bigl(P^\perp+\frac{\lambda+i\kappa}{\lambda-i\kappa}P\Bigr),$$
for $\kappa$ real. Then
$$
\big<\vartheta(\phi)\phi^{-1},h\big>\ =\ \frac{1}{2\pi i}\  {\rm Trace}
\oint_\gamma \
\frac{-2i\kappa Ps_1 \ \lambda^n}{(\lambda-i\kappa)(\lambda+i\kappa)}\
d\lambda\ =\
-(i\kappa)^{n}\big(1+(-1)^{n-1}\big)\ {\rm Trace}(Ps_1)\ ,
$$
which is zero if $n$ is even, and if $n$ is odd we calculate
$$
f_c\ =\ -4\kappa^ni(-1)^{(n-1)/2}\Big[
{\rm Trace}(Ps_1)\Big]_{x=-\infty}^\infty\ =\ -4\kappa^ni(-1)^{(n-1)/2} \Big[
\frac{i}{4}\ \frac{\bar\mu+\mu}{1+\mu\bar\mu}\Big]_{x=-\infty}^\infty\ ,
$$
and inserting the $x$ dependence of $\mu$ shows that
$$
f_c\ =\ -2\ |\kappa|^n\ (-1)^{(n-1)/2}\ .
$$

The calculation for the breather is more complicated. In this case
$\phi=\psi  e^{\beta u s_3}$, where $\psi$ is given by (\ref{eq: brpsi}).
We write $\psi=\psi_1\psi_2$, where
$$
\psi_1\ =\
\Big(P_{ 1}^\perp\ +\ P_{ 1}\
\frac{\lambda-\bar\alpha}{\lambda-\alpha}\Big) \quad{\rm and}\quad
\psi_2\ =\
\Big(P_{ 2}^\perp\ +\ P_{ 2}\ \frac{\lambda+\bar\alpha}{\lambda+\alpha}\Big)\ .
$$
Now $\vartheta (\phi)\phi^{-1}=\vartheta
(\psi_1)\psi_1^{-1}+\psi_1\vartheta (\psi_2)\psi_2^{-1}\psi_1^{-1}$,
and we can calculate
$$
\big<\vartheta(\psi_1)\psi_1^{-1},h\big>\ =\ \frac{1}{2\pi i}\  {\rm Trace}
\oint_\gamma \
\frac{-(\alpha-\bar\alpha) P_1s_1 \
\lambda^n}{(\lambda-\alpha)(\lambda-\bar\alpha)}\ d\lambda\ =\
-(\alpha^n-\bar\alpha^n)\ {\rm Trace}(P_1s_1)\ .
$$
The term $\big<\psi_1\vartheta (\psi_2)\psi_2^{-1}\psi_1^{-1},h\big>$
has the same limit as $x\to\pm\infty$ as the simpler term
$\big<\vartheta (\psi_2)\psi_2^{-1},h\big>$, because $\psi_1$ tends to a
limiting loop which commutes with
$s_1$. Now we can write
$$
f_c\ =\ 2\Big[\big<\vartheta
(\phi)\phi^{-1},h\big>\Big]_{x=-\infty}^\infty\ =\ -2
(\alpha^n-\bar\alpha^n)
\Big[{\rm Trace}(P_1s_1)+
(-1)^n {\rm Trace}(P_2s_1)\Big]_{x=-\infty}^\infty\ .
$$
For the breather case we can assume that $\alpha$ lies in the upper half
plane, in which case the
equation above gives zero for $n$ even, and for $n$ odd,
$$
f_c\ =\ 2i
(\alpha^n-\bar\alpha^n)\ .
$$

The integer $n$ of the $n^{\rm th}$ conserved charge is the Lorentz spin.
The fact that these charges were zero for $n$ even was observed beforehand
in \cite{OT}, in the more general context of the
affine Toda field theories,
since the elements of the principal Heisenberg subalgebra
of the Lie algebra of $\cal{M}$,
only has odd principal grades. In \cite{OT} it was shown that elements of
the principal Heisenberg subalgebra generated the higher conserved charges,
with the
derivation operator in the principle grade acting as the Lorentz boost, indeed
the principal Heisenberg subalgebra is nothing more than the subalgebra of
the Lie algebra of $\cal M$ which commutes with $J$ and $K$.
 It is then an immediate consequence that the Lorentz
spins of the charges, which are measured by the values of the principal grade,
are only non-zero for the exponents of the affine algebra $m$.
For $m=su(2)$, these are the odd integers. The argument presented here
is a concrete verification of this.

We would also like to bring to the reader's attention the expressions
given by the
central parts of $\phi^{-1}J\phi$, and $\phi^{-1}K\phi$,
$$<J,\vartheta(\phi)\phi^{-1}>\quad, {\rm and} \quad
<K,\vartheta(\phi)\phi^{-1}>,$$
respectively, discussed in
our previous paper \cite{bj1}, which were identified as the
energy and
momentum {\it densities} in light-cone co-ordinates integrated up to a point
$x$. It is evident that these are the same as the
Hamiltonian expressions (\ref{eq: Ham}), for $c=J$  and $c=K$,
provided we evaluate the
differences as
$x\rightarrow\pm\infty$ of the integrated
expressions.  This is because the Hamiltonian expressions only give  the
total quantities.

\resection{The splitting of the symplectic form.}
We must now address the practical concerns of calculating the symplectic form.
First we deal with the normalisation, that is $\phi=N\psi e^{\beta us_3}$,
where $\psi=1$
at $\lambda=\infty$, and where the constant matrix $N$
(which commutes with $s_1$) may be needed to satisfy the symmetry condition.
We wish to rewrite the expression for the
symplectic form
in terms of $\psi$, which is easier to deal with.  After doing this we see
that the form splits into two parts, one due to `self' interactions and
the other due to `mutual' interactions.

It is not too difficult to see that $N$ cancels immediately from
the expression for the symplectic form, as it cancels from $\phi^{-1} \phi_w$,
and does not affect $b$ at all. We shall therefore continue assuming that
$N=1$.

\noindent{\bf Proposition 9.1}

Consider the pole at $\alpha$, we write $\psi=\zeta\chi$,
where $\zeta$ is of the form (\ref{eq: the_form}) having a pole at $\alpha$,
and $\chi$ is regular at $\alpha$. Perform the usual factorisation
(\ref{eq: the_factorisation}) on $\zeta$ only, that is $a\zeta_0=\zeta d$,
this defines $d\in {\cal M}$. Then the $\alpha$ contribution to the
symplectic form (\ref{eq: the_result}) is
\beq
\omega(v,w)\big|_{\alpha}\ =\ \Big[\big<
 d\zeta_0^{-1}\zeta_{0v}d^{-1}-\zeta^{-1}\zeta_v,
 \zeta^{-1}\zeta_w\big>
\ -\ \big<
 \chi_{v}\chi^{-1},
 \zeta^{-1}\zeta_w\big>_{\alpha,\bar\alpha}\Big]_{x=-\infty}^\infty \ .
\label{eq: two_pieces}\eeq
The notation $<,>_{\alpha,\bar\alpha}$ means take the contributions
to the integral in the inner product only from the poles at
$\alpha$ and $\bar\alpha$.

We can
interpret the first term
here as a self interaction term of the $\alpha$ pole with itself, and the
second term as the mutual interaction
of the $\alpha$ pole with the other poles.
The total form is evidently the sum of such terms for each of the poles
$\alpha$.

\noindent{\bf Proof}

First note that
$$
\phi^{-1}\phi_w\ =\ e^{-\beta u s_3} \psi^{-1}\psi_w e^{\beta u s_3}\ +\
\beta u_w s_3\ ,
$$
and
\beq
b_v b^{-1}\ =\ \beta u_{0v}bs_3b^{-1}\ +\ e^{-\beta u s_3}
\big( \psi^{-1}a\psi_{0v}\psi_0^{-1}a^{-1}\psi-\psi^{-1}\psi_v\big)e^{\beta
u s_3}\ -\ \beta u_v s_3\ ,\label{eq: bvb} \eeq
Now look at the contribution to
$\big<b_vb^{-1}, \phi^{-1} \phi_w\big>$, which makes up the form
(\ref{eq: the_result}), from the $\beta u_{0v}bs_3b^{-1}$
term in (\ref{eq: bvb}),
which is
$$
\big< \beta u_{0v}bs_3b^{-1},\phi^{-1} \phi_w\big>\ =\
\beta u_{0v}\big< s_3,b^{-1}\phi^{-1} \phi_wb\big>\ =\
\beta u_{0v}\big< s_3,\phi_0^{-1} \phi_{0w}\big>\ -\ \beta u_{0v}\big<
s_3,b^{-1}b_w\big>\ .
$$
Here the second term is zero as the inner product vanishes on
analytic functions, and the first term cancels on taking the difference of
its values between
$x=R$ and $x=-R$. The contribution to the inner product
$\big<b_vb^{-1}, \phi^{-1} \phi_w\big>$ from the $\beta u_v s_3$ term of
(\ref{eq: bvb})
tends to zero as $x\to\pm\infty$, and we are left with the contribution
$$
\big<
e^{-\beta u s_3}
\big( \psi^{-1}a\psi_{0v}\psi_0^{-1}a^{-1}\psi-\psi^{-1}\psi_v\big)e^{\beta
u s_3},
e^{-\beta u s_3} \psi^{-1}\psi_w e^{\beta u s_3}\ +\ \beta u_w s_3\big>\ ,
$$
which simplifies to
$$
\big<
 \psi^{-1}a\psi_{0v}\psi_0^{-1}a^{-1}\psi-\psi^{-1}\psi_v,
 \psi^{-1}\psi_w \ +\ \beta u_w s_3\big>\ ,
$$
and the part containing $\beta u_w s_3$ vanishes as it is the inner product
of two analytic functions,
giving the simpler result
$$
\big<
 \psi^{-1}a\psi_{0v}\psi_0^{-1}a^{-1}\psi-\psi^{-1}\psi_v,
 \psi^{-1}\psi_w\big>\ =\
\big<
 c\psi_0^{-1}\psi_{0v}c^{-1}-\psi^{-1}\psi_v,
 \psi^{-1}\psi_w\big>
\ .
$$
Here we have used the factorisation $a\psi_0=\psi c$. As
$ c\psi_0^{-1}\psi_{0v}c^{-1}-\psi^{-1}\psi_v$ is analytic, we only have to
sum over the residues
of $\psi^{-1}\psi_w$ to calculate the inner product.

Let us look at the contribution from the poles at $\alpha$ (and
$\bar\alpha$) to the inner product. We
suppose that $\psi=\zeta\chi$, where $\zeta$ has a pole at $\alpha$, and $\chi$
is regular at $\alpha$. Then we can write
$$
\psi^{-1}\psi_w\ =\ \chi^{-1}\chi_w\ +\ \chi^{-1}\zeta^{-1}\zeta_w\chi\ .
$$
If we also write $a\zeta_0=\zeta d$, then $c=\chi^{-1}d\chi_0$, and the
$\alpha$
and $\bar \alpha$ contribution to the
inner product is
\begin{eqnarray}
\big<
 c\psi_0^{-1}\psi_{0v}c^{-1}-\psi^{-1}\psi_v,
 \psi^{-1}\psi_w\big>_{\alpha,\bar\alpha} & = & \big<
 c\psi_0^{-1}\psi_{0v}c^{-1}-\psi^{-1}\psi_v,
 \chi^{-1}\zeta^{-1}\zeta_w\chi\big>_{\alpha,\bar\alpha}  \cr\cr
& =  & \big<
 d\chi_{0v}\chi_0^{-1}d^{-1},
 \zeta^{-1}\zeta_w\big>_{\alpha,\bar\alpha}\ -\ \big<
 \chi_{v}\chi^{-1},
 \zeta^{-1}\zeta_w\big>_{\alpha,\bar\alpha}  \cr\cr
& & \quad +\
\big<
 d\zeta_0^{-1}\zeta_{0v}d^{-1}-\zeta^{-1}\zeta_v,
 \zeta^{-1}\zeta_w\big>_{\alpha,\bar\alpha}\label{eq: expr1}
\end{eqnarray}
The first term in (\ref{eq: expr1}) is
\begin{eqnarray*}
 \big<
 \chi_{0v}\chi_0^{-1},
 d^{-1}\zeta^{-1}\zeta_w d\big>_{\alpha,\bar\alpha} & = &
\big<
 \chi_{0v}\chi_0^{-1},
 \zeta_0^{-1}\zeta_{0w}\big>_{\alpha,\bar\alpha} \ -\
\big<
 \chi_{0v}\chi_0^{-1},
 d^{-1}d_w\big>_{\alpha,\bar\alpha}\ ,
\end{eqnarray*}
the last term of this vanishes as there is no pole at
$\alpha$ or $\bar\alpha$, and the first term vanishes on taking the difference
between the two asymptotic values of $x$.
We can now write the $\alpha$-contribution to
the inner product as
$$
\big<
 d\zeta_0^{-1}\zeta_{0v}d^{-1}-\zeta^{-1}\zeta_v,
 \zeta^{-1}\zeta_w\big>_{\alpha,\bar\alpha}
\ -\ \big<
 \chi_{v}\chi^{-1},
 \zeta^{-1}\zeta_w\big>_{\alpha,\bar\alpha} \ ,
$$
which can be rewritten as
$$
\big<
 d\zeta_0^{-1}\zeta_{0v}d^{-1}-\zeta^{-1}\zeta_v,
 \zeta^{-1}\zeta_w\big>
\ -\ \big<
 \chi_{v}\chi^{-1},
 \zeta^{-1}\zeta_w\big>_{\alpha,\bar\alpha} \ ,
$$
as $\zeta^{-1}\zeta_w$ only has poles at $\alpha$ and $\bar\alpha$.
$\quad\square$

\resection{The mutual interaction term.}

\noindent {\bf The limiting behaviour of the meromorphic loops.}

\noindent {\bf Lemma $10.1$}

The projections $P_i$, making up the product of meromorphic loops of the
form (\ref{eq: the_form})  for $\phi(\lambda)$, commute with $s_1$ in the
limits $x\rightarrow\pm\infty$.

\noindent {\bf Proof}

\noindent From the equations of motion of the linear system
$$\phi^{-1}\partial_+\phi = A - \phi^{-1} J \phi,\quad
{\rm and}\quad\phi^{-1}\partial_-\phi= B - \phi^{-1} K \phi.$$
in the limits $x\rightarrow\pm\infty$, (for finite $t$), the left-hand sides
of these equations are zero, and $A\rightarrow J$,  $B\rightarrow K$,
so we conclude that $\phi(\lambda, \pm\infty)$ commutes with $s_1$.
We add a soliton
to the system by multiplying a previous $\phi$ by a meromorphic unitary loop
on the right
of the form (\ref{eq: the_form}), and adjusting the normalisation.
The space-time behaviour of the previous
$\phi$ is unchanged. We also find that the limit of the
resultant $\phi$ commutes with $s_1$,
so we see that the limit of the meromorphic loop we have added
also commutes with $s_1$.
By taking residues therefore any of the $P_i$ in the products of the form
(\ref{eq: the_form}) making up $\phi$ also commute with $s_1$.
 $\quad\square$

For the following, recall that all matrices commuting with $s_1$
commute amongst themselves.

\noindent {\bf Lemma $10.2$}

We factor $\psi$ into the form $\psi\ =\ \zeta_\beta \chi_\beta$,
where $\zeta_\beta$ has only a pole at $\beta$, and $\chi_\beta$
 is regular at $\beta$. $\zeta_\beta$ is of the form (\ref{eq: the_form}),
where we denote the relevant projection by $P_\beta(x)$.
Then for any two poles $\beta_1$
and $\beta_2$ in $\psi$
$$
{\rm Trace}\big(P_{\beta_1}(\infty)P_{\beta_2}(\infty)-P_{\beta_1}(-\infty)
P_{\beta_2}(-\infty)\big)=0\ .$$

\noindent{\bf Proof}

This result is almost immediate from the form (\ref{eq: onesol}) of the
projection for the one-soliton solution. In the limit
($x\rightarrow-\infty$, $\Im(\beta)>0$) or ($x\rightarrow\infty$,
$\Im(\beta)<0$),
$$P_\beta\rightarrow\frac12\pmatrix{1&1\cr 1& 1},$$
and for
($x\rightarrow\infty$, $\Im(\beta)>0$) or ($x\rightarrow-\infty$,
$\Im(\beta)<0$),
$$P_\beta\rightarrow\frac12\pmatrix{1&-1\cr -1& 1}.$$
A simple calculation establishes the validity of the statement.$\quad\square$

\noindent{\bf Proposition $10.3$}

The mutual interaction term, the second term
of (\ref{eq: two_pieces}), vanishes, i.e.
$$\Big[\bigl<\chi_{\alpha v}\chi_{\alpha}^{-1},
 \zeta_{\alpha}^{-1}\zeta_{\alpha
w}\bigr>_{\alpha,\bar\alpha}\Big]_{x=-\infty}^\infty=0\ .$$

\noindent{\bf Proof}

Given a unitary meromorphic loop $\psi$ with
$\psi(\infty)=1$, we can factor it into
\beq
\psi\ =\ \zeta_\beta \chi_\beta\ ,\label{eq: zetachi}
\eeq
where $\zeta_\beta$ has only a pole at $\beta$, and $\chi_\beta$
 is regular at $\beta$, but will contain a pole at $\alpha$, c.f. equation
(\ref{eq: two_pieces}),  unless we choose $\beta=\alpha$.
In the generic case where poles are simple,
\begin{eqnarray}
\zeta_\beta & =& P_\beta^\perp\ +\
\frac{\lambda-\bar\beta}{\lambda-\beta}P_\beta\quad,\quad
\zeta^{-1}_\beta\  =\ P_\beta^\perp\ +\
\frac{\lambda-\beta}{\lambda-\bar\beta}P_\beta
\ , \cr\cr
\zeta_\beta^{-1}\zeta_{\beta v} & =&
\frac{\beta-\bar\beta}{\lambda-\beta}P_{\beta v}P_\beta \ +\
\frac{\beta-\bar\beta}{\lambda-\bar\beta}P_\beta P_{\beta v} \ +\
\frac{\lambda(\beta_v-\bar\beta_v)+ \bar\beta_v\beta-\beta_v\bar\beta }
{(\lambda-\beta)(\lambda-\bar\beta)}P_\beta\ ,\label{eq: zeta}
\end{eqnarray}
where $\beta_v$ and $P_{\beta v}$ are the changes in $\beta$ and $P_\beta$,
respectively, due to the
vector $v_0$. In deriving this formula we must remember that $P^\perp
P_v=P_vP$, which
comes from differentiating the equation $P^2=P$.
From (\ref{eq: zetachi}), we compute
\beq
\psi^{-1}\psi_v\ =\ \chi_\beta^{-1}\zeta_\beta^{-1}\zeta_{\beta
v}\chi_\beta\ +\
\chi_\beta^{-1}\chi_{\beta v}\ ,\label{eq: psiipsi}
\eeq
and noting that $\psi^{-1}\psi_v$ is a meromorphic function, zero at infinity,
and has simple poles at $\beta$ and $\bar\beta$,  we can write
$\psi^{-1}\psi_v$ in terms of its simple poles as
$$
\psi^{-1}\psi_v =  \sum_\beta \Big( \frac{{\rm
res}_\beta(\psi^{-1}\psi_v)}{\lambda-\beta}
\ +\ \frac{{\rm
res}_{\bar\beta}(\psi^{-1}\psi_v)}{\lambda-\bar\beta}\Big).  $$
From (\ref{eq: psiipsi}) this is equal to
\beq \psi^{-1}\psi_v =  \sum_\beta \Big(
\frac{\chi_\beta(\beta)^{-1}\ {\rm
res}_\beta(\zeta_\beta^{-1}\zeta_{\beta v})\ \chi_\beta(\beta)}
{\lambda-\beta}
\ +\
\frac{\chi_\beta(\bar\beta)^{-1}\ {\rm
res}_{\bar\beta}(\zeta_\beta^{-1}\zeta_{\beta v})
\ \chi_\beta(\bar\beta)}
{\lambda-\bar\beta}\Big)\label{eq: psiipsi2}
\eeq
We re-arrange formula (\ref{eq: psiipsi}):
\beq
\chi_{\alpha v}\chi_\alpha^{-1}\ =\ \chi_\alpha\psi^{-1}\psi_v
\chi_\alpha^{-1}\ -\
\zeta_\alpha^{-1}\zeta_{\alpha v}\ ,
\label{eq: re-arrange}\eeq
and calculate $\chi_{\alpha v}\chi_\alpha^{-1}(\alpha)$, which we will
eventually insert
into  $<\chi_{\alpha v}\chi_{\alpha}^{-1},
 \zeta_{\alpha}^{-1}\zeta_{\alpha w}>_{\alpha,\bar\alpha}$.
Insert the form (\ref{eq: psiipsi2}) into (\ref{eq: re-arrange}),
and consider the contribution to the sum from the $\beta\neq\alpha$ poles.

\noindent From Eqn. (\ref{eq: zeta}), and noting that
$P_{\beta v}\rightarrow 0$, as $x\rightarrow\pm\infty$,
${\rm res}_\beta(\zeta_\beta^{-1}\zeta_{\beta v})\to \beta_v
P_\beta(\pm\infty)$,
and ${\rm res}_{\bar\beta}(\zeta_\beta^{-1}\zeta_{\beta v})\to
-\bar\beta_v P_\beta(\pm\infty)$,
as $x\to\pm\infty$, and since
$\chi_\alpha(\alpha)\chi_\beta(\beta)^{-1}$, and
$\chi_\alpha(\alpha)\chi_\beta(\bar\beta)^{-1}$
tend to matrices
which commute with $s_1$, c.f. Lemma $10.1$,
and therefore commute respectively with
${\rm res}_\beta(\zeta_\beta^{-1}\zeta_{\beta v})$,
and ${\rm res}_{\bar\beta}(\zeta_\beta^{-1}\zeta_{\beta v})$
 after the limits
$x\rightarrow\pm\infty$,
we get the contribution to $\chi_{\alpha v}\chi_\alpha^{-1}(\alpha)$ of
\beq
\sum_{\beta\neq\alpha} \Big( \frac{\beta_v }{\alpha-\beta}
\ -\ \frac{\bar\beta_v }{\alpha-\bar\beta}\Big)P_\beta(\pm\infty) \ .
\label{eq: contrib}\eeq
The contribution to $\chi_{\alpha v}\chi_\alpha^{-1}(\alpha)$ from the
$\bar\alpha$
pole is zero, and for the $\alpha$ pole we get the limit as
$\lambda\rightarrow\alpha$ of
$$
\frac{\chi_\alpha(\lambda)
\chi_\alpha(\alpha)^{-1}\ {\rm
res}_\alpha(\zeta_\alpha^{-1}\zeta_{\alpha v})\ \chi_\alpha(\alpha)
\chi_\alpha(\lambda)^{-1}- {\rm
res}_\alpha(\zeta_\alpha^{-1}\zeta_{\alpha v})
}{\lambda-\alpha}.
$$
This is
$$
\big[\chi'_\alpha(\lambda)
\chi_\alpha(\alpha)^{-1}\ ,\  {\rm
res}_\alpha(\zeta_\alpha^{-1}\zeta_{\alpha v})\big]\ +\ O(
\lambda-\alpha)\ ,
$$
which tends to zero on putting $\lambda=\alpha$ and taking limits as
$x\to\pm\infty$, since after these limits, both terms in the commutator
separately commute with $s_1$, c.f. Lemma $10.1$,
and so commute with each other.
Thus $\chi_{\alpha v}\chi_\alpha^{-1}(\alpha)$ is equal to formula
(\ref{eq: contrib}), at the limits when $x\rightarrow\pm\infty$.
In the formula (\ref{eq: zeta}) for  $\zeta_\alpha^{-1}\zeta_{\alpha
w}(\lambda)$
we note that $P_{\beta v}\rightarrow 0$, as $x\rightarrow\pm\infty$, so
that only the last term of
(\ref{eq: zeta}) contributes. Since $\chi_{\alpha
v}\chi_\alpha^{-1}(\lambda)$  is
regular at $\alpha$,
 we see that the $\alpha$ pole contribution to
the inner product
$\big<
 \chi_{\alpha v}\chi_\alpha^{-1},
 \zeta_\alpha^{-1}\zeta_{\alpha w}\big>_{\alpha,\bar\alpha}$ is
made up of sums of terms proportional to
$$
{\rm
trace}\big(P_\beta(\infty)P_\alpha(\infty)-P_\beta(-\infty)P_\alpha(-\infty)\big
)\,
$$
for $\beta\neq\alpha$.
From Lemma $10.2$, these terms are all zero.
Likewise the contribution from the $\bar\alpha$ pole gives zero, so we
conclude that
the mutual interaction term $\big<
 \chi_{\alpha v}\chi_\alpha^{-1},
 \zeta_\alpha^{-1}\zeta_{\alpha w}\big>_{\alpha,\bar\alpha}$
 contributes nothing to the symplectic form.$\quad\square$

\resection{The self interaction term.}

\noindent{\bf Proposition 11.1}

Consider the self-interaction term, $<
 d\zeta_0^{-1}\zeta_{0v}d^{-1}-\zeta^{-1}\zeta_v,
 \zeta^{-1}\zeta_w>$,  the first term
of (\ref{eq: two_pieces}), then we have
\beq\big<
 d\zeta_0^{-1}\zeta_{0v}d^{-1}-\zeta^{-1}\zeta_v,
 \zeta^{-1}\zeta_w\big>\ =\ \pm 2{\rm Re\ }\frac{\alpha_v Q_{w}-\alpha_w
Q_{v}}{\alpha Q}\ ,
\eeq
(where $+$ corresponds to $\Im \alpha>0$ and $-$ to $\Im \alpha<0$),
or in wedge notation,
\beq
=\ \pm 2{\rm Re\ }\frac{d\alpha\wedge dQ}{\alpha Q}.\label{eq: form2}\eeq
Here $Q$ is the coefficient specifying the initial projection of the
left-most ordered meromorphic loop in the product of loops. For the two-soliton
case, comparing with Eqn. (\ref{eq: Q1Q2}), this $Q$ is either
$Q_1$ or $Q_2$ depending on whether
the pole at $i\kappa_1$ or $i\kappa_2$ is considered. The total form is the sum
of the contributions of this type.

\noindent{\bf Proof}

We take
$$
\zeta\ =\ P^\perp\ +\ P\frac{\lambda-\bar\alpha}{\lambda-\alpha}\ ,
$$
which gives, as before,
\beq
\zeta^{-1}\zeta_{ v} \ =\
\frac{\alpha-\bar\alpha}{\lambda-\alpha}P_{ v}P \ +\
\frac{\alpha-\bar\alpha}{\lambda-\bar\alpha}P P_{ v} \ +\
\frac{\lambda(\alpha_v-\bar\alpha_v)+ \bar\alpha_v\alpha-\alpha_v\bar\alpha }
{(\lambda-\alpha)(\lambda-\bar\alpha)}P\ ,\label{eq: zeta2}
\eeq
and for convenience we shall write
$n=\lambda(\alpha_v-\bar\alpha_v)+ \bar\alpha_v\alpha-\alpha_v\bar\alpha$.
We also need to look at the value of $d=\zeta^{-1}a\zeta_0$ at $\alpha$ and
$\bar\alpha$.
We recall that $\zeta^{-1}=(P^\perp+{\lambda-\alpha\over\lambda-\bar\alpha}P)$,
and then
\beq
d(\lambda) =
\big(P^\perp+{\lambda-\alpha\over\lambda-\bar\alpha}P\big)a(\lambda)
\big(P_0^\perp+{\lambda-\bar\alpha\over\lambda-\alpha} P_0\big)
\ .\label{eq: whatisd}
\eeq
Now, from (\ref{eq: zeta2}), we find that the first entry in the inner
product to be calculated is
\begin{eqnarray*}
H(\lambda) & = & d\zeta_0^{-1}\zeta_{0v}d^{-1}-\zeta^{-1}\zeta_v \\
& = &
\frac{\alpha-\bar\alpha}{\lambda-\alpha}(dP_{0v}P_0d^{-1}-P_vP) \ +\
\frac{\alpha-\bar\alpha}{\lambda-\bar\alpha}(dP_0P_{0v}d^{-1}-PP_v) \ +\
\frac{n (dP_0d^{-1}-P) }
{(\lambda-\alpha)(\lambda-\bar\alpha)}\ .
\end{eqnarray*}
The contribution to the total inner product from the $\alpha$ and
$\bar\alpha$ poles, noting that $P^2=P$, using cyclicity of the trace, and
$P^\perp P_w=P_wP$, is:
\beq
{\rm Trace}\Big(P(\frac{\alpha_w}{\alpha}
H(\alpha)-\frac{\bar\alpha_w}{\bar\alpha} H(\bar\alpha))P\Big)\ +\
(\alpha-\bar\alpha)\ {\rm Trace}\Big((\frac{PH(\alpha)P^\perp}{\alpha}
+\frac{P^\perp H(\bar\alpha) P}{\bar\alpha} )P_w\Big)\ ,
\label{eq: total_contrib}\eeq
and we calculate, by substituting (\ref{eq: whatisd}) for $d(\lambda)$, and
writing $a$ for $a(\lambda)$,
\begin{eqnarray*}
PH(\lambda)P& = &
(\alpha-\bar\alpha)\Big(\frac{PaP_{0v}P_0a^{-1}P}{\lambda-\bar\alpha}+
\frac{PaP_0P_{0v}P_0^\perp a^{-1}P}{\lambda-\alpha}\Big)\ +\ n
\frac{PaP_0a^{-1}P-P}{(\lambda-\alpha)(\lambda-\bar\alpha)}   \\
PH(\lambda)P^\perp & = &
(\alpha-\bar\alpha)\Big(\frac{(\lambda-\alpha)PaP_{0v}P_0a^{-1}P^\perp}
{(\lambda-\bar\alpha)^2} +
\frac{PaP_0P_{0v}a^{-1}P^\perp-PP_v}{\lambda-\bar\alpha}\Big)  \ +\ n
\frac{PaP_0a^{-1}P^\perp}{(\lambda-\bar\alpha)^2}  \\
P^\perp H(\lambda) P & = &
(\alpha-\bar\alpha)\Big(\frac{P^\perp a P_{0v}P_0
a^{-1}P-P_vP}{\lambda-\alpha}  +
\frac{(\lambda-\bar\alpha)P^\perp a P_0
P_{0v}a^{-1}P}{(\lambda-\alpha)^2}\Big)   \ +\ n
\frac{P^\perp a P_0 a^{-1}P}{(\lambda-\alpha)^2}.
\end{eqnarray*}
We know that $P$ is the orthogonal projection to $a(\alpha){\rm im}P_0$, which
gives rise to the equations
\beq
Pa(\alpha)P_0\ =\ a(\alpha)P_0 \quad{\rm and}\quad P_0a(\alpha)^{-1}P\ =\
a(\alpha)^{-1}P\ .\label{eq: laws}
\eeq
Using these equations (\ref{eq: laws}), we find
\begin{eqnarray*}
PH(\alpha)P & = & PaP_{0v}P_0a^{-1}P\ -\
(\alpha-\bar\alpha)PaP_0P_{0v}a^{-1}a_\lambda a^{-1} P  \\
& &\ +\
\alpha_v \big(Pa_\lambda P_0a^{-1}P-PaP_0a^{-1}a_\lambda a^{-1}P\big).
\end{eqnarray*}
If we take the equation $Pa(\alpha)P_0a(\alpha)^{-1}P=P$, derived from
(\ref{eq: laws}), and differentiate it,
we find that
$$
P_vaP_{0}a^{-1}P+PaP_{0v}a^{-1}P+PaP_{0}a^{-1}P_v+
\alpha_v(Pa_\lambda P_{0}a^{-1}P-PaP_{0}a^{-1}a_\lambda a^{-1}P)\ =\ P_v\ ,
$$
and using this we can write
\begin{eqnarray*}
{\rm Trace}\big(PH(\alpha)P\big) & = & -\ {\rm Trace}\big( aP_{0}a^{-1}P_v
\ +\  (\alpha-\bar\alpha)PaP_0P_{0v}a^{-1}a_\lambda a^{-1} P \big).
\end{eqnarray*}
We also find that
\begin{eqnarray*}
PH(\alpha)P^\perp & = & aP_0P_{0v}a^{-1}P^\perp\ -\ PP_v\ +\ \frac{\alpha_v}
{\alpha-\bar\alpha}aP_0a^{-1}P^\perp   \\
{\rm Trace}\big(PH(\alpha)P^\perp P_w\big) & = & {\rm
Trace}\big(PaP_0P_{0v}P_0^\perp a^{-1} P_w \ - \
PP_vP_w\ +\ \frac{\alpha_v}
{\alpha-\bar\alpha}aP_0a^{-1}P_w P\big).
\end{eqnarray*}
The combination from the $\lambda=\alpha$ terms of (\ref{eq: total_contrib})
is, writing $a$ for $a(\alpha)$,
$$
{\rm Trace}\big(\alpha_w PH(\alpha)P+(\alpha-\bar\alpha)PH(\alpha)P^\perp
P_w\big)/\alpha\ =\
\alpha^{-1} \times
$$
$$
{\rm Trace}\big( \alpha_v aP_0a^{-1}P_w P-\alpha_w a P_0 a^{-1}P_v P\ +
(\alpha-\bar\alpha)\big( PaP_0P_{0v}P_0^\perp a^{-1}P_w+
\alpha_w PaP_0P_{0v}P_0^\perp (a^{-1})_\lambda P-
PP_v P_w\big)\big).
$$
By differentiating the equation
$P_0^\perp a^{-1}(\alpha)P=0$, derived from (\ref{eq: laws}), we obtain
$$
-P_{0w} a^{-1}P\ +\ \alpha_w P_0^\perp (a^{-1})_\lambda P \ +\ P_0^\perp
a^{-1}P_w\ =\ 0\ ,
$$
so
$${\rm Trace}\big(\alpha_w PH(\alpha)P+(\alpha-\bar\alpha)PH(\alpha)P^\perp
P_w\big)/\alpha=\qquad\qquad\qquad \qquad\qquad\qquad$$
$$
\qquad\qquad\qquad{\rm Trace}\big( \alpha_v aP_0a^{-1}P_w P-\alpha_w a P_0
a^{-1}P_v P\ +
 (\alpha-\bar\alpha)\big( P_0P_{0v}P_{0w}-
PP_v P_w\big)\big)/\alpha\ .$$
The term $PP_v P_w$ tends to zero as $x\to\pm\infty$, and $P_0P_{0v}P_{0w}$
vanishes on taking the difference between two values of $x$, leaving the
$\lambda=\alpha$ terms from
(\ref{eq: total_contrib}) as
$$
\frac1\alpha\big[{\rm Trace}\big(\alpha_w
PH(\alpha)P+(\alpha-\bar\alpha)PH(\alpha)P^\perp P_w\big)
\big]^\infty_{x=-\infty}=\qquad\qquad\qquad$$
\beq
\qquad\qquad\qquad\ \frac1\alpha\big[{\rm Trace}\big( \alpha_v
aP_0a^{-1}P_w P-\alpha_w a P_0 a^{-1}P_v
P\big)
 \big]^\infty_{x=-\infty}\ .\label{eq: leaving}
\eeq
Let $P$ be the orthogonal projection to the complex vector $(1,\mu)$. Then
$$
P\ =\ \frac{1}{1+\mu\bar\mu}
\left(\matrix{ 1 & \bar \mu \cr \mu & \mu\bar\mu}  \right)
\quad {\rm and}\quad
P_v P\ =\ \frac{\mu_v}{(1+\mu\bar\mu)^2}
\left(\matrix{ -\bar\mu & -\bar \mu^2 \cr 1 & \bar\mu }  \right).
$$
Since $a(\lambda)=e^{-Jx_+-Kx_-}$,
it can be written in the form
$$
a(\alpha)\ =\ \frac 12\left(\matrix{ r+1/r & r-1/r \cr r-1/r & r+1/r}
\right)\ ,
$$
where
$
r=\exp\big(-imx(\alpha -\alpha^{-1})/2\big)
$.
Here we have set $t=0$, although the limits derived below are equally
valid for any finite $t$.
Then it is possible to calculate
\begin{eqnarray*}
{\rm Trace}\big( \alpha_vaP_0a^{-1}P_w P\big)\ & = & -
\frac{(\mu_0^2-1)(\bar\mu_0^2-1)(|r|^4-1)}{(1+|\mu_0|^2)\big(
(1-\mu_0)(1-\bar\mu_0)+|r|^4(1+\mu_0)(1+\bar\mu_0)\big)}\
\alpha_v\frac{r_w}{r}  \\
 &  & +\
\frac{(\bar\mu_0^2-1)(|r|^4-1)}{(1+|\mu_0|^2)\big(
(1-\mu_0)(1-\bar\mu_0)+|r|^4(1+\mu_0)(1+\bar\mu_0)\big)}\ \alpha_v\mu_{0w}
\end{eqnarray*}
As $\alpha_vr_w/r=-imx(1+\alpha^{-2})\alpha_v\alpha_w/2$,
the first term vanishes on
subtraction of the term with
$v$ and $w$ reversed. If ${\rm Im}({\alpha})>0$, so that
the lower limit corresponds to $|r|=0$,
equation (\ref{eq: leaving}) becomes
\beq
\frac1\alpha\big[{\rm Trace}\big( \alpha_v aP_0a^{-1}P_w P-\alpha_w a P_0
a^{-1}P_v P\big)
\big]^\infty_{|r|=0}\ =\
\frac{2(\alpha_v\mu_{0w}-\alpha_w\mu_{0v})}{\alpha(\mu_0^2-1)}\ .
\label{eq: above}\eeq
The overall sign of (\ref{eq: above}) is reversed if ${\rm Im}(\alpha)<0$.
If we use the coordinate $Q$, c.f. Eqn (\ref{eq: Q}), where
$$
\mu_0\ =\ \frac{1-iQ}{1+iQ}\ ,
$$
we can rewrite Eqn. (\ref{eq: above}) as
$$
\frac1\alpha\big[{\rm Trace}\big( \alpha_v aP_0a^{-1}P_w P-\alpha_w a P_0
a^{-1}P_v P\big)
 \big]^\infty_{|r|=0}\ =\ \frac{\alpha_v Q_{w}-\alpha_w Q_{v}}{\alpha Q}.
$$
We also calculate the $\lambda=\bar\alpha$ parts of (\ref{eq: total_contrib}),
and find the total contribution to the symplectic form
$$\big<
 d\zeta_0^{-1}\zeta_{0v}d^{-1}-\zeta^{-1}\zeta_v,
 \zeta^{-1}\zeta_w\big>\ =\ \pm
2{\rm Re\ }\frac{\alpha_v Q_{w}-\alpha_w Q_{v}}{\alpha Q}\ ,
 $$
where $+$ corresponds to ${\rm Im}(\alpha)>0$ and $-$ to ${\rm Im}(\alpha)
<0$.\quad$\square$

\resection{Final statement of result.}

\noindent{\bf The two-soliton case.}
By combining Propositions $9.1$, $10.3$ and $11.1$, and by considering
the contributions from the poles at $i\kappa_1$, and $i\kappa_2$,
we have derived the symplectic form $\omega$ for the two-soliton solution
$$\frac\omega{2}={d\kappa_1\wedge dQ'_1\over \kappa_1 Q'_1}
+ {d\kappa_2\wedge dQ'_2\over \kappa_2 Q'_2}.$$
Throughout this section, we must take the real parts of the expressions
giving the forms. Also, for the solitons,
without loss of generality we can take $\kappa>0$, and therefore take
the plus sign of (\ref{eq: form2}).
If $\kappa<0$, the minus sign in (\ref{eq: form2}) is to be expected
because of the relation of the position $x_0$ of the soliton to $Q$, see
Eqn. (\ref{eq: x_0}). If we wished to write the form in terms
of the rapidity $\theta$ and position of the soliton, we must substitute
$|\kappa|=e^{-\theta}$, and therefore pick up a minus sign if $\kappa<0$,
this cancels with the minus appearing in (\ref{eq: form2}), so in both cases
we would have an overall plus sign.

Here the explicit two-soliton solution in these co-ordinates is written as
\beq e^{-i{\beta u\over 2}}=
{(1+iY^{-1}Q'_1W_{1}-iY^{-1}Q'_2W_{2}+Q'_1Q'_2W_{1}W_{2})
\over (1-iY^{-1}Q'_1W_{1}+iY^{-1}Q'_2W_{2}+Q'_1Q'_2W_{1}W_{2})}, \label{eq:
intermediate}\eeq
with
$$Y={\kappa_2-\kappa_1\over\kappa_2+\kappa_1},$$
compare with Eqn. (\ref{eq: soweget}), and following.  Also,
$W_i=e^{m(\kappa_ix_+-\kappa_i^{-1}x_-)}$, and $\kappa_i$ is related
to the rapidity $\theta_i$ of the soliton by $e^{-\theta_i}=|\kappa_i|$.

If we write (\ref{eq: intermediate})  in terms of the co-ordinates
$Q_1, Q_2$ which are
more familiar to us, so that the two-soliton solution is
$$e^{-i{\beta u\over 2}}=
{(1+iQ_1W_{1}-iQ_2W_{2}+Y^{2}Q_1Q_2W_{1}W_{2})
\over (1-iQ_1W_{1}+iQ_2W_{2}+Y^{2}Q_1Q_2W_{1}W_{2})}, $$
by defining
$$Q_1=Y^{-1}Q'_1,\qquad\qquad Q_2=Y^{-1}Q'_2.$$
Then the symplectic form is
\bea \frac\omega{2}&=&{d\kappa_1\wedge d(YQ_1)\over \kappa_1 YQ_1}
+ {d\kappa_2\wedge d(YQ_2)\over \kappa_2 YQ_2}\cr\cr
&=&{d\kappa_1\wedge dQ_1\over  \kappa_1 Q_1}
+ {d\kappa_2\wedge dQ_2\over  \kappa_2 Q_2}
+{d\kappa_1\wedge dY\over \kappa_1 Y}
+{d\kappa_2\wedge dY\over \kappa_2 Y}.\eea
However, it is easy to show that
$${d\kappa_1\wedge dY\over\kappa_1}={d\kappa_2\wedge dY\over\kappa_2}=
{2\over{(\kappa_1+\kappa_2})^2}d\kappa_1\wedge d\kappa_2$$
so we can rewrite the form as
$$\frac\omega{2} = {d\kappa_1\wedge d(Y^2Q_1)\over \kappa_1 Y^2Q_1}
+ {d\kappa_2\wedge dQ_2\over\kappa_2 Q_2}$$
or
$$\frac\omega{2} ={d\kappa_1\wedge dQ_1 \over \kappa_1 Q_1}
+ {d\kappa_2\wedge d(Y^2Q_2)\over \kappa_2 Y^2Q_2},$$
these last two expressions are ones familiar from Babelon and Bernard
\cite{BB}, where the form is represented in terms of  ``in'' and
``out'' co-ordinates, respectively. The multiplication of one of the $Q$'s by
$Y^2$,
is precisely the shift in one of the co-ordinates of the solitons by the
standard time delay, proportional to  $\log{(Y^2)}$, for sine-Gordon.

\noindent{\bf The $n$-soliton case.} The form is evidently the sum of
single soliton contributions given by the left-most ordered $Q$'s,
denoted $Q'_i$. However, it must be remembered that these $Q'_i$ are not
the co-ordinates which appear
in the standard $n$-soliton solution, that is the $n$-soliton version
of (\ref{eq: two_sol_sol}), or the $n$-soliton solutions
given in \cite{BB}, (in \cite{BB}, they are denoted by $a_i$).
We denote these co-ordinates $Q_i$.
 The two sets of co-ordinates are related
by $$Q'_i=Q_i\prod_{j\neq i}{\kappa_j-\kappa_i\over\kappa_i+\kappa_j}.$$
Exactly as we have seen for the two-soliton case above, we can then
make contact with the diagonal result of \cite{BB}, where the form is
given in terms of `in' or `out' co-ordinates.

\noindent{\bf The breather case.}  Begin with the factorisation
(\ref{eq: brpsi}), where the left ordered projections are $P_1$ and $P_3$.
Without loss of generality we may assume that the imaginary part of
$\alpha$ is positive.
Then the contribution to the symplectic form from the $\alpha$ pole is
$$
{\omega\over 2}\Bigl|_{\alpha}=\frac{d\alpha\wedge dQ_{1}}
{\alpha Q_1}\ .
$$
To find the contribution from the $-\alpha$ pole we use equation
 (\ref{eq: newsymmetry2}) and (\ref{eq: V3}) to find $Q_3=1/Q_1$,
and we can now give the total
symplectic form for a breather, remembering
to take the real part of this,  as
$$
{\omega\over 2}\Bigl|_{\alpha,-\alpha}=2\frac{d\alpha\wedge dQ_{1}}
{\alpha Q_1}\ .
$$
The $Q$ used here is the one appearing in the left-most ordered factor,
the factor with a pole at $\alpha$,
in a product of loops representing many breathers and solitons.
It is crucial to remember that this $Q$ will  be shifted, as compared
with the $Q$ in the single breather solution (\ref{eq: classical_breather}),
by factors which arise when the other breathers and solitons
to the right of the left-most
ordered factors are taken into account. Compare with the $n$-soliton case.

\noindent{\bf Acknowledgements}

This work was started when P.R.J was at the Dept. of Physics, University
of Wales Swansea, supported by PPARC. It was completed whilst P.R.J
was supported by the European Union TMR network programme, contract
ERBFMRXCT960012.
 
  \end{document}